\documentclass[conference,compsoc]{IEEEtran}

\usepackage{amsthm}
\usepackage{comment}
\usepackage{threeparttable}
\usepackage{multirow}
\theoremstyle{definition}

\makeatletter
\def\thm@space@setup{\thm@preskip=1.5pt
\thm@postskip=1.5pt}
\makeatother

\usepackage{graphicx}
\usepackage{algorithm}
\usepackage{amssymb}
\usepackage{algpseudocode}

\usepackage{tikz}
\usepackage{amsfonts}
\usepackage{ifthen}

\def\BibTeX{{\rm B\kern-.05em{\sc i\kern-.025em b}\kern-.08em
    T\kern-.1667em\lower.7ex\hbox{E}\kern-.125emX}}
    
\usepackage{subcaption}
\usepackage{gensymb}
\usepackage{pifont}

\newboolean{commentsOn}
\setboolean{commentsOn}{true}

\usepackage{xspace}

\DeclareRobustCommand*\circled[1]{\tikz[baseline=(char.base)]{ \node[shape=circle,draw,color=white,fill=black,inner sep=0.5pt] (char){#1};}}

\def\ie{{i.e.},~}
\def\eg{{e.g.},~}

\def\etal{{et al.}~}

\usepackage{microtype}

\newcommand{\shortsectionBf}[1]{\vspace{2pt}
\noindent {\bf #1}}

\usepackage[shortlabels]{enumitem}

\usepackage{amsmath}
\mathchardef\mhyphen="2D

\usepackage{setspace}

\usepackage[utf8]{inputenc}
\usepackage[T1]{fontenc}
\usepackage{textcomp}

\usepackage{array}

\usepackage{hyperref}
\mathchardef\mhyphen="2D

\definecolor{lightblue}{rgb}{0.54, 0.81, 0.94}
\newcommand{\takeawaybox}[1]{\vspace{2pt}
\noindent {\setlength{\fboxsep}{0pt}\colorbox{lightblue}{\bf #1 }}}

\newcommand\blfootnote[1]{%
  \begingroup
  \renewcommand\thefootnote{}\footnote{#1}%
  \addtocounter{footnote}{-1}%
  \endgroup
}

\usepackage{url}

\begin{document}
\title{Understanding Users' Security and Privacy Concerns and Attitudes Towards\\ Conversational AI Platforms}

\author{
    \IEEEauthorblockN{Mutahar Ali}
    \IEEEauthorblockA{University of California, Irvine\\
     mutahara@uci.edu}

     \and

    \IEEEauthorblockN{Arjun Arunasalam}
    \IEEEauthorblockA{Purdue University\\
    aarunasa@purdue.edu}
    
    \and
    
    \IEEEauthorblockN{Habiba Farrukh}
    \IEEEauthorblockA{University of California, Irvine,\\
    habibaf@uci.edu}
}

\maketitle
\IEEEpeerreviewmaketitle

\begin{abstract}
The widespread adoption of conversational AI platforms has introduced new security and privacy risks. While these risks and their mitigation strategies have been extensively researched from a technical perspective, users' perceptions of these platforms' security and privacy remain largely unexplored. In this paper, we conduct a large-scale analysis of over 2.5M user posts from the r/ChatGPT Reddit community to understand users' security and privacy concerns and attitudes toward conversational AI platforms. Our qualitative analysis reveals that users are concerned about each stage of the data lifecycle (\ie collection, usage, and retention). They seek mitigations for security vulnerabilities, compliance with privacy regulations, and greater transparency and control in data handling. We also find that users exhibit varied behaviors and preferences when interacting with these platforms. Some users proactively safeguard their data and adjust privacy settings, while others prioritize convenience over privacy risks, dismissing privacy concerns in favor of benefits, or feel resigned to inevitable data sharing. Through qualitative content and regression analysis, we discover that users' concerns evolve over time with the evolving AI landscape and are influenced by technological developments and major events. Based on our findings, we provide recommendations for users, platforms, enterprises, and policymakers to enhance transparency, improve data controls, and increase user trust and adoption.
\end{abstract}

\section{Introduction}
\label{sec:intro}
\blfootnote{
\noindent
\copyright~2025 IEEE. Personal use of this material is permitted.  Permission from IEEE must be obtained for all other uses, in any current or future media, including reprinting/republishing this material for advertising or promotional purposes, creating new collective works, for resale or redistribution to servers or lists, or reuse of any copyrighted component of this work in other works.

Published in the IEEE Symposium on Security and Privacy (S\&P), 2025. DOI: \url{https://doi.ieeecomputersociety.org/10.1109/SP61157.2025.00241}
}Large language models (LLMs)~\cite{brown2020language, chowdhery2023palm} have revolutionized artificial intelligence (AI), driving the widespread adoption of conversational AI platforms (\eg ChatGPT \cite{openai2024chatgpt}, Gemini \cite{google2024gemini}, and Claude \cite{anthropic2024claude}). These platforms allow users to interact with LLM-based AI systems through different modalities, such as text and voice. They are being integrated into a diverse range of applications, including customer service~\cite{pandya2023automating, rajaraman2024customer_service, technologyreview2024conversationalAI}, personal assistants~\cite{google_gemini_assistant, apple_intelligence}, healthcare~\cite{forbes2024healthcare, clusmann2023future}, and finance~\cite{wu2023bloomberggpt, li2023large}, due to their advanced natural language abilities, enabling more intuitive user interactions.

However, as conversational AI platforms become more popular, concerns surrounding their security and privacy (S\&P) have also intensified. These platforms present unique S\&P risks as their human-like conversational abilities can inadvertently encourage users to share sensitive information more freely than with traditional interfaces~\cite{ischen2020privacy}. Prior work has also demonstrated that LLMs can memorize and reproduce sensitive information from their training data in response to malicious prompts~\cite{carlini2021extracting, staab2024beyond}.

High-profile data leaks \cite{Browne2023}, corporate restrictions on platform usage due to suspected sensitive data exposures~\cite{ray2023samsung, bloomberg2023citigroup, businessinsider2023amazon}, and the drive for regulations (\eg the White House Executive Order on AI adoption~\cite{whitehouseExecutiveOrder}) further highlight the critical nature of these risks.

While prior research has predominantly focused on the technical aspects of securing LLMs~\cite{jang2022knowledge, kandpal2022deduplicating, yu2021differentially, lison-etal-2021-anonymisation}, and in turn these platforms, there is a notable gap in understanding how users perceive and respond to these S\&P risks. Preliminary studies have leveraged mixed methods (\eg surveys, semi-structured interviews) to investigate users' S\&P concerns about conversational AI platforms~\cite{zhang2024s, gumusel2024user}. These studies provide valuable initial insights into privacy harms and risks associated with AI systems, including common types of personal information users disclose and factors that influence trust in these platforms. However, these studies are conducted in controlled contexts with limited participant diversity and capture user concerns at a single point in time. Consequently, a comprehensive understanding of users' S\&P concerns, behaviors, and preferences in interactions with conversational AI platforms across diverse user bases and how they change over time is needed.

To address this gap, we complement prior works by conducting a large-scale analysis of real-world, organic discussions on Reddit~\cite{reddit} to investigate the following research questions:

\begin{itemize}
    \item \textbf{RQ1} What are users' S\&P concerns related to conversational AI platforms?
    \item \textbf{RQ2} What are users' S\&P attitudes toward conversational AI platforms?
    \item \textbf{RQ3} How do users' S\&P concerns evolve over time, and how do major events in the AI ecosystem influence users' S\&P concerns and attitudes? 
\end{itemize}

To answer these questions, we collect and analyze a dataset comprising $\sim2.5$M posts ($\sim180$K submissions and $\sim2.35$M comments) from \texttt{r/ChatGPT}, the largest subreddit dedicated to conversational AI platforms, with over $7.4$M members. Using keyword filtering and stratified sampling, we select and manually annotate a subset of $1,200$ posts from the entire corpus to fine-tune a RoBERTa-based binary classifier~\cite{liu2019roberta} for identifying S\&P-related content. Applying this classifier to the entire dataset yields $30{,}240$ S\&P-related posts. We iteratively sample and code a subset of these posts through qualitative analysis to develop a codebook. We then leverage an LLM-based multi-class classifier for post-hoc data labeling, assigning thematic codes to the entire S\&P dataset. Finally, we conduct an interrupted time series regression analysis~\cite{mcdowall2019interrupted} to investigate how users' S\&P concerns evolve over time, specifically in response to major events in the AI ecosystem.

Our qualitative and quantitative analysis reveals that users are primarily concerned about what personal and proprietary data these platforms collect and why ($43.7\%$). They also worry about how the collected data is used ($22.5\%$), particularly for training models or third-party sharing, and its retention ($9.5\%$). Users also seek assurance in platform security and security of apps integrating conversational AI platforms ($28.9\%$), compliance with privacy laws ($9.6\%$), and desire transparent data handling practices and clear control over their data ($11.1\%$).

We observe that users exhibit varied S\&P behaviors and preferences when interacting with conversational AI platforms. Some are proactive in safeguarding their data, adjusting privacy settings, and limiting the information they share. Others are inquisitive, seeking information about how their data is collected, used, and stored. Some prioritize convenience and the benefits these platforms offer, often downplaying or dismissing privacy risks. Meanwhile, others feel resigned to data sharing as an unavoidable aspect of modern digital interactions. 

Our longitudinal analysis further reveals that users' S\&P concerns evolve over time, influenced by key events such as platform updates, regulatory changes, feature releases, and security incidents. For example, Microsoft's investment in OpenAI~\cite{ForbesMicrosoftChatGPT2023} sparked increased concern about third-party sharing, while Italy's temporary ban on ChatGPT~\cite{nytimes2023chatgptitalyban} due to GDPR violations intensified discussions about compliance with data protection regulations.

Our study extends efforts in understanding users' S\&P concerns and attitudes towards conversational AI platforms. It also highlights how these concerns shift in response to significant industry and regulatory events in the AI ecosystem. We conclude by synthesizing recommendations for key stakeholders, including recommendations for users to take proactive steps in managing their data privacy, platforms to improve transparency and create more intuitive data controls, enterprises to define clear usage guidelines, and policymakers to establish clear regulations and promote the standardization of privacy information. These recommendations collectively contribute to a safer and more trustworthy AI ecosystem.

In this paper, we make the following contributions:
\begin{itemize}
    \item We conduct a large-scale analysis of more than $2.5$M user posts from Reddit to understand users' S\&P concerns and attitudes towards conversational AI platforms.
    \item We investigate how users' concerns evolve over time and identify the impact of major industry and regulatory events on users' concerns through regression analysis.
    \item We provide actionable recommendations for key stakeholders to support the safe, transparent, and responsible deployment and use of conversational AI platforms.
\end{itemize}

\section{Related Work}

\shortsectionBf{Users' S\&P Concerns and Attitudes.}
Prior research has explored users' security and privacy (S\&P) concerns~\cite{kumaraguru2005privacy}, behaviors~\cite{kokolakis2017privacy}, preferences~\cite{lin2014modeling,naeini2017privacy}, and attitudes~\cite{dupree2016privacy}. Studies have also explored users' S\&P attitudes and concerns toward specific platforms and technologies. For example, studies on smart homes and IoT devices have highlighted user apprehensions about data collection and device security~\cite{SmartHome, zeng2017end, haney2020user, tabassum2019don, zeng2019understanding}. Similarly, studies on web browsers and online social networks have investigated users' views and satisfaction with privacy settings, uncovering a range of concerns about data sharing and control~\cite{hossain2015privacy}. While prior research has examined general S\&P attitudes or concerns toward specific technologies, we focus on the unique challenges and concerns that arise from the human-like interactions facilitated by conversational AI.

\shortsectionBf{Online Discussions on S\&P.}
Online discussion platforms (\eg Reddit, X, StackOverflow) offer spaces for open dialogue, mutual learning, and peer support, providing rich, real-world data that reflects user concerns and behaviors. Researchers have leveraged online discussions to study S\&P from different perspectives~\cite{al2023sentiment, wei2020twitter, tahaei2020understanding, vetrivel2023examining, SmartHome, DevelopersPrivacyReddit, IBSA, horawalavithana2019mentions}. For example, Li~\etal~\cite{SmartHome} analyzed Reddit posts to investigate users' S\&P considerations when adopting smart home technologies. Similarly, Tahaei~\etal\cite{tahaei2020understanding} conducted a qualitative analysis of privacy-related discussions on Stack Overflow to explore privacy challenges faced by developers. Wei~\etal~\cite{wei2020twitter} studied user perceptions of targeted advertising by analyzing Twitter data. Several works have also investigated user privacy feedback by analyzing app reviews on app stores (\eg Google Play Store)~\cite {Hark, AndroidAppReviewSP24, mukherjee2020empiricalstudyuserreviews, UserPerspectivesOnAppPrivacyICSE, ShortTextLargeEffect}. These studies illustrate the potential of online discourse to surface concerns and behaviors not easily captured through traditional surveys or interviews. Our work builds on this approach by analyzing discussions on the \texttt{r/ChatGPT} subreddit to capture nuanced and evolving user S\&P perspectives unique to conversational AI platforms.

\shortsectionBf{S\&P in Conversational AI Platforms.}
Large Language Models (LLMs) \cite{brown2020language, chowdhery2023palm} are increasingly being leveraged to power AI systems that engage users in natural language across multiple modalities (\eg text and voice). We refer to these AI systems as conversational AI platforms. These platforms vary in complexity, ranging from basic chatbots to intelligent agents with task completion capabilities, and have unique S\&P challenges. LLMs' human-like conversational abilities can encourage users to share sensitive information more freely than with traditional interfaces~\cite{ischen2020privacy}. Moreover, prior works show that LLMs can memorize and reproduce sensitive information from their training data~\cite{carlini2021extracting}, raising concerns about data leakage and privacy breaches.

A large body of work has explored the S\&P of conversational AI platforms from a technical perspective~\cite{carlini2021extracting, staab2024beyond, kandpal2022deduplicating, yu2021differentially, lison-etal-2021-anonymisation, liu2024formalizing}. These studies primarily aim to understand vulnerabilities inherent to LLM architectures and behaviors. However, far less attention has been paid to the human factors associated with these risks. 

Recent work by Lee~\etal~\cite{lee2024don} studied AI practitioners' awareness, motivation, and ability to mitigate AI-related privacy risks, revealing limited awareness of AI-specific threats and insufficient incentives for privacy work. A few studies have also explored users' S\&P concerns towards conversational AI platforms. Gumusel~\etal~\cite{gumusel2024user} conducted semi-structured interviews with $13$ participants to investigate privacy harms and risks in conversational AI. Zhang~\etal~\cite{zhang2024s} used a mixed-methods approach, analyzing a dataset~\cite{ShareGPT52K} of user interactions with ChatGPT to categorize data disclosure scenarios in conversational AI and conducting interviews with $19$ participants to investigate privacy concerns and disclosure behaviors. They identified patterns in user behavior and factors influencing trust during interactions.

While these studies provide valuable initial insights, they have three main limitations. First, they primarily involve participants from the United States, which may not represent the global user base of conversational AI platforms. Privacy concerns and regulatory landscapes vary across jurisdictions (\eg GDPR~\cite{gdpr} in Europe), leading to differing user concerns and behaviors. Second, these works focus on end-user interactions, overlooking S\&P concerns from developers and enterprises deploying these platforms. Lastly, they explore users' perspectives in a controlled context at a single point in time, failing to capture how user concerns and attitudes evolve in response to new features and emerging risks.

In contrast, our work analyzes a large-scale dataset from Reddit, capturing a broader range of user concerns and attitudes over time. By focusing on discussions from a diverse user base, we capture real-world user interactions and include perspectives influenced by regional privacy regulations like the GDPR. Our analysis spans a significant time frame, allowing us to observe how users' S\&P concerns evolve in response to major events such as new features, policy changes, and regulatory actions. This dynamic perspective offers a more comprehensive understanding of user attitudes toward S\&P in conversational AI platforms.

\section{Methodology}
\begin{figure}[t]
    \centering
    \includegraphics[width=\linewidth]{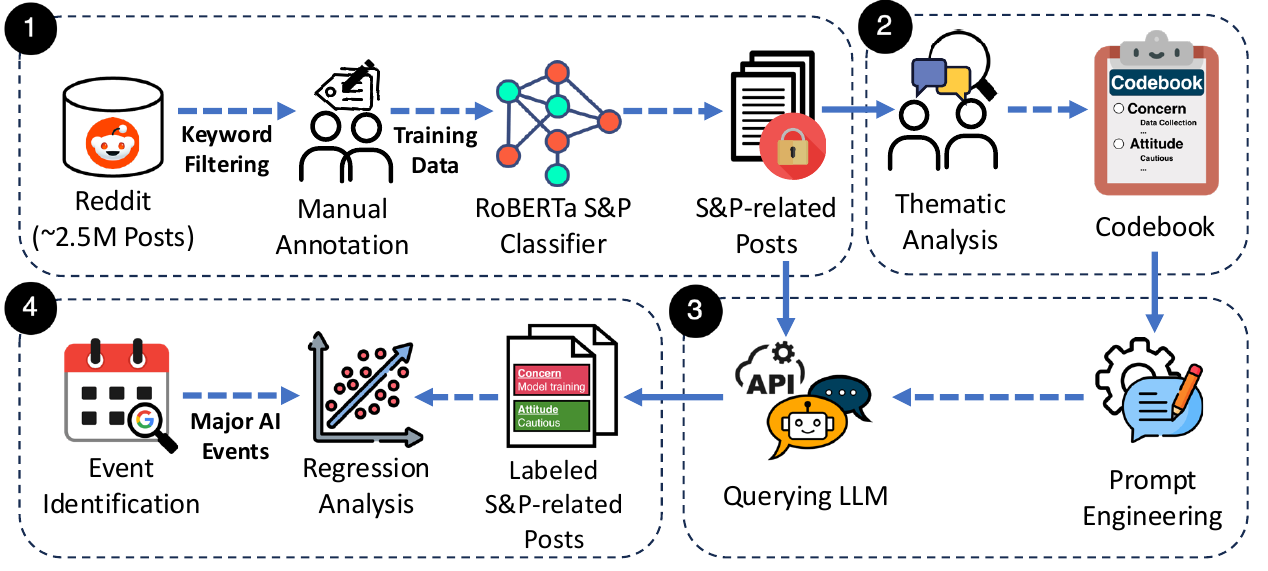}
    \caption{Methodology overview.}
    \label{fig:methodology}
\end{figure}

\label{sec:methods}

To understand users' security and privacy (S\&P) concerns and attitudes toward conversational AI platforms, we conduct a multi-stage analysis of discussions on Reddit~\cite{reddit}. Reddit is an online discussion forum organized into topic-specific communities called subreddits, where users share, discuss, and vote on content in threaded structures. Each thread consists of a submission (the original post) and a series of comments. Reddit discussions represent large-scale, organic user-generated data 
that offer valuable insights into users' concerns, perceptions, behaviors, and information-seeking practices in the real-world~\cite{SmartHome, DevelopersPrivacyReddit, IBSA}.

Figure~\ref{fig:methodology} presents an overview of our methodology, which involves four main steps. We begin by collecting a large dataset of Reddit posts and annotating a subsample to train a classifier that identifies content related to S\&P~\circled{1}. Next, we conduct a qualitative analysis to identify users' S\&P concerns and attitudes towards conversational AI platforms~\circled{2}. Building on this, we leverage an LLM-based multi-class classifier to label all S\&P-related posts with thematic codes derived from our qualitative analysis~\circled{3}. Finally, we perform a regression analysis to investigate how users' S\&P concerns shift in response to major events in the AI ecosystem~\circled{4}.

\subsection{Identifying S\&P-related posts}
\label{subsec:sp_classification}

To understand users' concerns and attitudes, we first compiled a list of popular subreddits related to conversational AI platforms using Reddit's search engine~\cite{reddit}, as shown in Appendix~\ref{appendix:subreddits} Table~\ref{tab:subreddits}. From this list, we selected \texttt{r/ChatGPT} for our analysis based on several reasons. First, it is the largest and most active subreddit focused on conversational AI with $7.4$M members - more than $4$ times the size of the next largest subreddit (\texttt{r/OpenAI}) - and ranks $86th$ among all Reddit communities. Second, it has high user engagement and activity ($\sim300$ daily posts and $\sim4{,}000$ comments). Third, despite its branding, \texttt{r/ChatGPT} regularly features discussions that extend beyond ChatGPT. For example, over $7{,}500$ posts mention local LLMs (\eg LLaMA~\cite{MetaLlama2_2023}), and more than $43{,}000$ reference other commercial platforms like Gemini~\cite{google2024gemini} and Claude~\cite{anthropic2024claude}), indicating a broader topical scope. Lastly, our qualitative analysis shows that the subreddit hosts diverse user attitudes (Section~\ref{sec:attitudes}), ranging from cautious and curious to privacy-dismissive and resigned. Therefore, selecting \texttt{r/ChatGPT} allows us to capture substantial data covering diverse S\&P concerns across different conversational AI platforms.  

To conduct our analysis, we gathered all posts from \texttt{r/ChatGPT} from its inception in December $2022$ till July $2024$, using publicly available Reddit dumps~\cite{pushshift}. After removing deleted posts, bot-generated content~\footnote{Bot-generated posts are identified using known bot signatures, such as the presence of the word ``bot'' in the username or phrases like ``Beep boop, I'm a bot''~\cite{hurtado2019bot}.}, and duplicate entries, our final dataset included $\sim2.5$M posts ($\sim180$K submissions and $\sim2.35$M comments, collectively termed ``posts'').

Given the dataset size, identifying S\&P-related discussion threads is a significant challenge. An intuitive approach is to perform a search with keywords associated with S\&P to surface relevant posts~\cite{DevelopersPrivacyReddit}. However, this approach can suffer from low coverage, missing relevant posts that use alternative phrasing, and high false positives, especially in communities like \texttt{r/ChatGPT} where users frequently share generative content (e.g., fictional stories, essays, or poems) that may include keywords in unrelated contexts. To address these limitations, we adopt a hybrid approach combining keyword filtering with machine learning.

\shortsectionBf{Keyword Filtering and Sampling.}
We began by reviewing the literature on S\&P issues in conversational AI and AI risk taxonomies \cite{carlini2021extracting, zhang2024s, gumusel2024user, yan2024protecting, yao2024survey, cui2024risk, yang2024memorization, carliniquantifying, hartmann2023sok} to compile an initial list of candidate keywords. We refined this list iteratively by searching Reddit posts, reviewing matched results, identifying new keywords, and updating the list accordingly. Using this iterative approach, we compiled a list of $118$ S\&P-related keywords and expressions, presented in Appendix~\ref{appendix:keywords}~Table~\ref{tab:keywords}.

To ensure balanced representation and reduce bias toward frequently mentioned keywords, we grouped our keywords into $6$ thematic categories based on prior works~\cite{cui2024risk,emami2021privacy,emami2019exploring}. We then performed stratified random sampling, selecting $100$ posts from each group, resulting in $600$ posts potentially related to S\&P. To enable the classifier to distinguish S\&P-related content from other topics and to capture themes outside our keyword list, we complemented this sample with $600$ randomly selected posts that did not contain any keywords. In total, our seed corpus consisted of $1,200$ posts, balanced between submissions and comments.

\shortsectionBf{Manual Annotation.}
To construct a reliable seed corpus for training our classifier, we manually annotated the $1{,}200$ sampled posts, labeling each as either S\&P-related or not S\&P-related. To ensure consistency and minimize subjectivity, two authors (experts in information S\&P) jointly reviewed an initial subset of $200$ posts to create a detailed annotation guide, defining clear boundaries between S\&P and non-S\&P content. Using this guide, the two authors then independently labeled the corpus, achieving high inter-rater agreement (Cohen's Kappa, $\kappa=0.82$). The final labeled dataset included $209$ S\&P-related posts ($17.4\%$) and $991$ non-S\&P posts.

\shortsectionBf{S\&P-Related Post Classification.}
To identify S\&P-related posts within our dataset, we fine-tuned a binary classifier using RoBERTa~\cite{liu2019roberta}, a pre-trained language model for text classification tasks. Given the nuanced nature of S\&P topics and the requirement for both high precision and recall, we selected RoBERTa based on its superior performance over other models (\ie DeBERTa~\cite{he2020deberta}, GPT-4~\cite{gpt4o} and Gemini Flash 1.5~\cite{gemini_flash}), as shown in Appendix~\ref{appendix:sp_classification}~Table~\ref{tab:sp_classification}. The classifier achieved an accuracy of $96\%$ and an F1-score of $82\%$. Applying this classifier to the entire dataset ($\sim 2.5$M posts), yielded $30{,}240$ S\&P-related posts ($1.2\%$), authored by $18{,}851$ unique Reddit users. Appendix~\ref{appendix:sp_posts_over_time} Figure~\ref{fig:sp_posts_over_time} shows the weekly counts of S\&P-related posts over time.

\subsection{Data Analysis}
\label{subsec:data_analysis}

\subsubsection{Sampling and Coding}
\label{subsubsec:sampling_and_coding}

To identify users' S\&P concerns and attitudes, we conducted a thematic analysis. We first developed an initial codebook informed by our keyword exploration and manual annotation for S\&P classification (Section~\ref{subsec:sp_classification}). To refine the codebook, two authors iteratively sampled and coded S\&P-related posts over two weeks until reaching thematic saturation~\cite{saunders2018saturation} at $440$ posts.

These $440$ posts were authored by $433$ unique Reddit users. With each sample, we updated our codebook by introducing new codes for emerging concerns and attitudes. We then grouped similar codes into broader themes that align with specific stages of the data lifecycle (data collection, usage, and retention) as well as broader concerns that span multiple stages (security vulnerabilities, regulatory compliance, and transparency and control). We do not present intercoder agreement as all posts were reviewed together~\cite{mcdonald2019reliability}.

\subsubsection{Prevalence Analysis}

To analyze the prevalence of concerns and their evolution across events (Section~\ref{sec:longitudinal_analysis}), we labeled the entire dataset of S\&P-related posts ($30{,}240$ posts) using an LLM-based multi-class classifier. Given the scale of our dataset, manually annotating all posts was infeasible. Therefore, similar to prior works \cite{IBSA}, we leveraged LLMs for post-hoc data labeling to label each post with codes from our qualitative analysis (Section~\ref{sec:concerns}). Specifically, we used GPT-4o~\cite{gpt4o} with a hierarchical prompting~\cite{liu2023hierarchical} approach. First, the model predicted the high-level concerns (themes) \eg data collection, data usage. Then, for each predicted theme, we used a separate prompt to predict the low-level concerns (sub-themes) \eg personal data, model training. We evaluated the classifier on our dataset of $440$ manually coded posts (Section~\ref{subsubsec:sampling_and_coding}), achieving an average accuracy of $96.8$\% and F1-score of $91.2$\%. Detailed performance metrics for each theme and sub-theme are presented in Appendix~\ref{appendix:multi-class}~Table~\ref{tab:multiclass_classification_results}. After applying the classifier to the entire dataset, we conducted a secondary evaluation to assess its performance on unseen data. We randomly sampled $100$ LLM-classified posts and manually coded them to generate ground truth labels. The classifier achieved an accuracy of $98.5$\% and an F1-score of $95.1$\% in labeling these posts.

\subsection{Ethical considerations}
We strictly follow ethical guidelines for research involving publicly available online data. All data used in this study was collected solely for non-commercial academic purposes. Before conducting this research, we consulted with the Institutional Review Board (IRB) at our institution. The study was deemed to be IRB exempt, as it involved the use of publicly available information.

To protect user privacy, all usernames have been anonymized, and no personal identifiers have been retained in our dataset. The data is used exclusively for internal analysis and has not been redistributed. Additionally, we do not release any derivative products, such as our classifier for identifying S\&P-related posts, to ensure the privacy of the individuals represented in the data.
\section{RQ1: Users' S\&P Concerns}
\label{sec:concerns}

\begin{figure}[t]
    \centering
    \includegraphics[width=\linewidth]{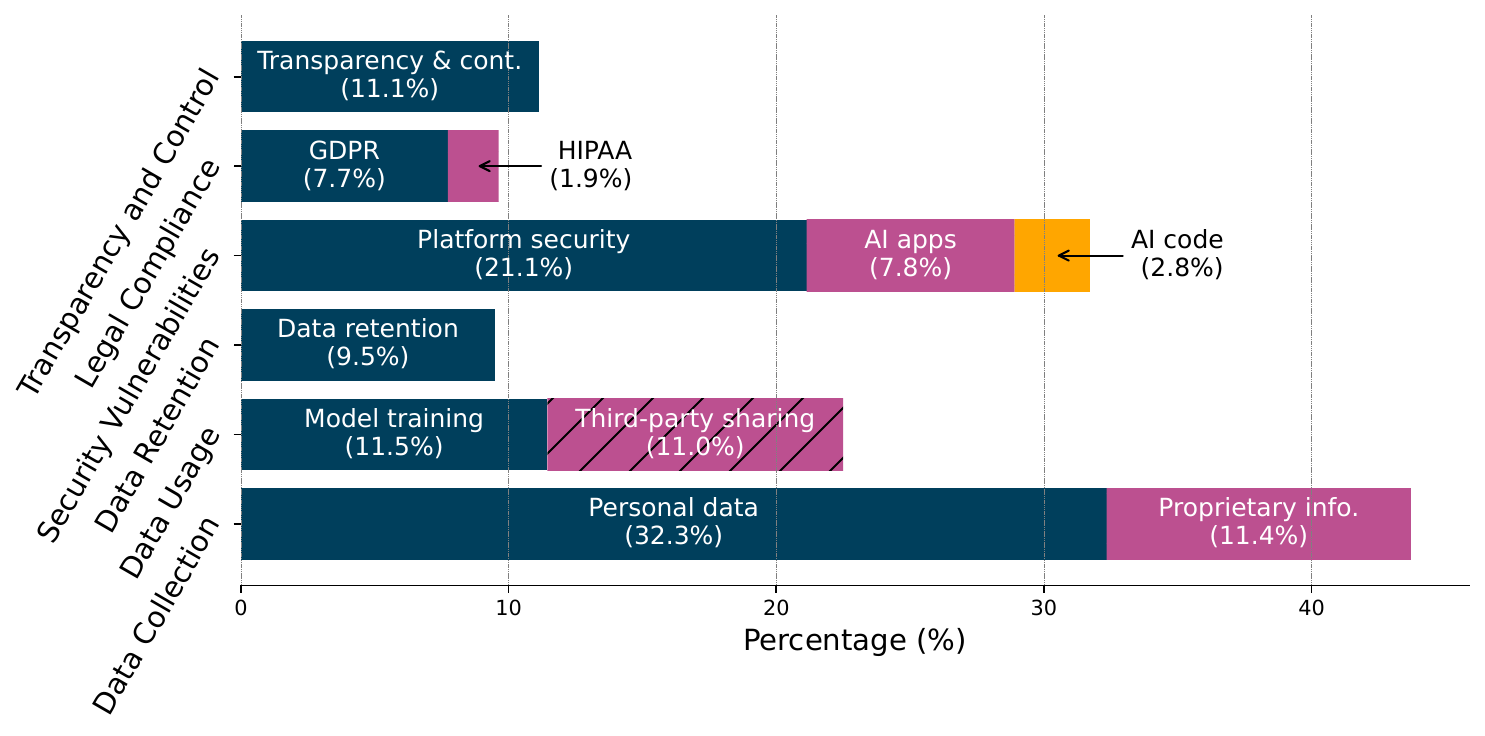}
    \caption{Prevalence of users' S\&P concerns.}
    \label{fig:prevalenceOfConcerns}
\end{figure}

Our analysis reveals a range of recurring security and privacy (S\&P) concerns regarding conversational AI platforms (\eg ChatGPT). We categorize these concerns into six main themes: ($1$) data collection, ($2$) data usage, ($3$) data retention, ($4$) security vulnerabilities, ($5$) legal compliance, and ($6$) transparency and control. Figure~\ref{fig:prevalenceOfConcerns} presents the prevalence of each theme and its sub-themes. These prevalence metrics can guide platforms in prioritizing improvements by highlighting the most common user concerns \eg the need for greater transparency and control around data collection and model training. Below, we detail our findings for each theme.

\subsection{Data Collection}

Users are wary of the nature and scope of data collected by conversational AI platforms. We discover concerns related to \textit{personal data} $-$ information that identifies individuals or reveals private details, and \textit{proprietary information} $-$ sensitive corporate information and intellectual property, e.g., business plans, trade secrets, ideas, and work products.

\subsubsection{Personal Data}

Users discuss that interacting with conversational AI platforms often requires sharing personal data, either directly or via passive data collection.

\shortsectionBf{Personally Identifiable Information (PII).}
PII refers to any data that can uniquely identify an individual~\cite{piidef}, such as \textit{``IP address, email, username, phone number, and device ID''}. Users express concern regarding PII collection by conversational AI platforms. For instance, one user expressed discomfort about the AI platform having access to their name and university information: \textit{``The data gathering is scary, it know[s] my name and my university''}. Some users review the privacy policies of popular AI platforms to understand their policies regarding PII collection. For example, one user pointed out that ChatGPT's privacy policy \textit{``[states] that they can personally identify you''}. Another user mentioned that \textit{``they can put the data and the identity info together''}.

\shortsectionBf{Activity and Behavioral Data.}
Users are concerned that conversational AI platforms may track and monitor various aspects of their digital behavior beyond conversational data. Their main concerns revolve around the collection of browsing activity, GPS location, and data from other apps on the user's devices. Users report instances where the AI response includes personal details that they did not provide, questioning whether the platform is monitoring their online activity and browsing habits. For example, one user noted, \textit{``Does ChatGPT have access to our browser \& site visits? It figured out my [university's] name despite [me] not previously mentioning it.''}

Users also worry about AI platforms accessing their location data without permission. For instance, one user noted ChatGPT's apparent knowledge of their location: \textit{``Does ChatGPT really have GPS location without enabling? I never told it my location before. It guessed down to the [right] town.''} In contrast, some users do not find this surprising, comparing it to how Google Maps works (\eg \textit{``Do you freak out when you Google `Food near me' and it actually searches restaurants in your area?''}).

\subsubsection{Proprietary Information}

Several discussions in our analysis involve employees in corporate settings expressing concern about using conversational AI platforms with confidential and proprietary information such as business plans, internal emails, trade secrets, project details, and software. Users also express concerns about the ownership and privacy of their intellectual property (IP).

\shortsectionBf{Corporate Data.}
Many users view the use of AI tools in professional environments as a \textit{``corporate nightmare''} due to the risk of exposure of sensitive data to unauthorized parties. Some users worry that this may lead to \textit{``consequences''} such as legal repercussions or suspension \eg \textit{``[ChatGPT] does not treat submitted info as confidential ... If anyone finds out you are using [it] in your day job ... you are probably in material breach of your employment contract.''}.

Many users are also wary of the privacy risks of using conversational AI with work-related information, especially in fields bound by confidentiality, such as legal, finance, and consulting roles. For example, one user expressed their concern regarding the use of AI tools in the legal profession: \textit{``The information you ask from [the AI] or give [it] access to is most likely not private. Isn't this a [major issue] in the law space [with] client confidentiality [being] broken?''}.

\shortsectionBf{Software.}
Many users share specific concerns about the privacy of code and software shared with platforms, expressing skepticism about the \textit{``settings to disable data collection''} and questioning how they can \textit{``prevent ChatGPT from spreading [their] information''} to keep sensitive code private. For instance, a user inquired, \textit{``When using GPT, is there a way to keep the code/data you are feeding it private, like if I am working with patented/proprietary code and do not want it exposed outside my company?''}

To address S\&P concerns, some conversational AI platforms offer specialized tiers for corporate use~\cite{google2024workspaceai,openai2024chatgptpricing} with added S\&P features (\eg exclusion of conversation data from model training). Nevertheless, users remain cautious and uncertain about the actual privacy protections these plans offer. For example, one user expressed concern about risks associated with ChatGPT Teams after reviewing OpenAI's privacy policy: \textit{``The data in Team tier is not used for training, but can still be viewed by other people … what's worse [is that] access is not limited to OpenAI staff''}. 

\shortsectionBf{Ideas and Creative Works.}
In addition to corporate concerns, users who turn to conversational AI platforms for creative tasks (\eg brainstorming, refining startup concepts) worry about these platforms \textit{``stealing''} or misappropriating their original ideas. For instance, one user questioned whether sharing a story draft with ChatGPT might result in the story being misused, asking, \textit{``If I had a story rough draft and I tell it to ChatGPT to get it refined, will the story end up getting stolen?''} Another user voiced suspicion that ChatGPT might be registering trademarks on their ideas, advising caution against using AI tools for creative tasks: \textit{``I'm very suspicious that ChatGPT is stealing my ideas [from] brainstorming and getting trademarks or copyrights on certain names or ideas before I get a chance to. Beware of using these tools for any personal projects or ideas or inventions.''}

\subsection{Data Usage}
\label{subsec:data_usage}

Concerns about data usage revolve around its use for improving AI models and sharing with third parties.

\subsubsection{Model Training}

We find that users have two primary concerns regarding the use of their data in these training processes: the involvement of human reviewers in evaluating user data for training and the potential for models to ``memorize'' and disclose specific details from user interactions.

\shortsectionBf{Human Review.}
Human oversight in model training frequently involves annotators reviewing user interactions to label data, guide improvements, and ensure accuracy \cite{achiam2023gpt}. Users express concerns that PII might be exposed during model training: \textit{``When an AI system is in beta testing, the inputs and outputs of the system may be read by humans who are helping to train the model. This means that your PII could potentially be seen by these humans.''}

We also find specific concerns related to popular conversational AI platforms. One Bard (now Gemini ~\cite{google2024gemini}) user advised against entering any private information in Bard, recalling Bard's explicit notice: \textit{``Your conversations are processed by human reviewers to improve the technologies powering Bard.''} Some users point out that ChatGPT lacks a similar warning, potentially leaving users unaware that human reviewers may access their interactions. For example, one user stated, \textit{``At least Google ``says'' not to [share private information]. While [its] competitor loves binge eating data.''} 

\shortsectionBf{Memorization.}
Prior work has shown that LLMs can \textit{``memorize''} and reproduce specific pieces of information from their training data \cite{carlini2021extracting}. Users worry that sensitive information entered into conversational AI platforms could inadvertently be disclosed to other platform users through the models' responses. For example, one user cautioned that \textit{``[ChatGPT] learns from the data''} while another noted that data can be \textit{``regurgitated to other users in some form.''} Users feel this is a major issue, as memorized data could include PII. For instance, one user described how ChatGPT \textit{``doxxed''} a random person in its response: \textit{``I sent ChatGPT a text about charisma I wanted him to summarize ... [it wrote] out [the] most personal and private information of a random man.''}

\subsubsection{Third-party Sharing}

Users share concerns about who may access their data, for what purpose, and under what conditions. For instance, one user asked, \textit{``Does OpenAI share user activity with third parties? And if so, which parties?''}

\shortsectionBf{Affiliates.}
Many users are skeptical of data sharing within corporate ecosystems, where affiliated companies might gain access to their information across interconnected platforms. This concern is highlighted by users of conversational AI platforms developed by large tech conglomerates. For instance, one user questioned whether Meta AI was \textit{``trained on Whatsapp chats''}, while another speculated that Microsoft CoPilot \textit{``already has [the] data''} from Outlook and Teams. Some users expressed concern regarding data sharing between Microsoft and OpenAI: \textit{``OpenAI [can] pump all that juicy data to Microsoft while tying it to your device.''}

\shortsectionBf{Data Brokers and Advertisers.}
Another major concern for users is the potential sale of their personal information to data brokers and its use for targeted advertising~\cite{juels2001targeted}. For example, one user expressed their discomfort after reading OpenAI's Terms of Service, stating that \textit{``[they allow] *all* personal information to be sold to any third party for any reason they think is OK ... specifically including advertisers and marketing services.''} Another user shared a similar worry: \textit{``[ChatGPT's privacy policy] literally says they will sell [users'] personal data.''}

Users foresee various privacy consequences of sharing data with brokers and advertisers. Some worry that targeted ads could manipulate their purchasing choices, while others express deeper concerns about personal data being shared with insurance companies or lenders, possibly affecting loan eligibility or insurance premiums. For instance, one user commented, \textit{``I'm more concerned about when advertisers get to influence training data. Google search is heinous enough at this point; imagine when it's even more subtle.''} Another user noted, \textit{``I would be more worried that my information [could] be sold to insurance brokers or lenders and that [may impact] my ability to get certain insurance or credit. Who would want to lend someone money for a house or give a suicidal person an insurance policy?''}

\shortsectionBf{Government Authorities.}
Conversational AI platforms may be legally obligated to disclose user data to government authorities in certain situations, such as when a conversation includes references to imminent harm or in response to a court order or subpoena~\cite{openai_privacy_policy}. This raises concerns among users about the privacy of their interactions with the platforms. For example, one user highlighted the ease of government access, noting, \textit{``If the US government subpoenas data you sent to OpenAI (including prompts and their responses) then OpenAI will almost certainly and immediately comply.''}

\subsection{Data Retention}
\label{subsec:data_retention}

User concerns around data retention mainly center around incomplete deletion, where data appears hidden rather than fully erased, and data storage practices.

\shortsectionBf{Incomplete Data Deletion.}
One common perception among users is that after deletion, data is removed from the user's view but persists on the backend. For instance, one user stated that \textit{``when you delete a conversation, ChatGPT actually hides it instead of deleting it.''} Another user raised suspicion, stating \textit{``I've said some stuff that I would rather not be seen by anyone. I've already click[ed] the delete button but is it really deleted?''} Commenting on OpenAI's Terms of Service (ToS), another user pointed out that \textit{``nothing in their ToS says they will remove content ... it's marked as deleted, and then hidden from the user.''} These posts reflect a broader sentiment of mistrust regarding data handling practices among conversational AI platforms. One user summarized this sentiment: \textit{``There is no way to know they will actually delete your data. If OpenAI was nonprofit, their claim would be more believable.''} 

\shortsectionBf{Data Storage Practices.}
Another user concern is that personal data is implicitly stored within model weights after training, making it nearly impossible to erase. One user noted: \textit{``OpenAI has used your personal data to train their model and cannot easily remove it.''} Some users are also concerned that this may violate data protection laws (\eg GDPR~\cite{gdpr}) under which \textit{``companies must let you delete your personal data''} (See Section~\ref{subsubsec:GDPR}). Moreover, users question whether platforms retain data for extended periods in ways that are inaccessible or unverifiable (\eg \textit{``We don't know if OpenAI archives the data or actually deletes it but my guess is they archive as much of it as they can.''})

\subsection{Security Vulnerabilities}

Apart from concerns regarding data collection, usage, and retention, users are also concerned about the security of conversational AI platforms and the data shared with them.

\subsubsection{Platform Security}

We find that users frequently express concerns about the security of conversational AI platforms, especially the risk that sensitive data might be exposed due to software bugs and data breaches.

\shortsectionBf{Software Bugs.}
Users worry that software bugs in AI platforms may have security implications, potentially exposing sensitive user data to unauthorized parties. A notable example occurred in early 2023 when a bug in ChatGPT allowed users to view the titles of other users' conversations \cite{Browne2023}, which one user described as \textit{“a massive privacy problem.”} Another user, alarmed by seeing unknown conversation titles in their history, decided to delete their account, fearing that their personal information was compromised: \textit{``I believe my information is shared by someone or mixed up with [another] person's information ... As soon as I am being refunded, I plan to delete my account''}. Some users believed that their account was \textit{``hacked''} after seeing unknown conversations in their chat history.

Users also specifically discuss the security of ChatGPT's code interpreter, fearing how vulnerabilities could be exploited to steal data. One user suggested that it can be used to steal user data by getting the user to \textit{``[paste] a malicious URL into ChatGPT''} and using it to \textit{``run instructions''} in the browser to \textit{``grab data and send it to a third-party server.''} Such concerns emphasize users' apprehension about the security robustness of these platforms.

\shortsectionBf{Data Breaches.}
Another significant user concern involves the risk of data breaches, which could expose private data from stored conversations. For instance, a user noted, \textit{``I'm scared that there's going to be a data breach and my information will be leaked.''} This apprehension makes users cautious about sharing personal details with AI platforms, as highlighted by a user who warned: \textit{``Be careful with the info you provide ChatGPT ... If the system is hacked or experiences a data breach, the PII you provide could be accessed and potentially misused by unauthorized parties.''} Another user explained how, even with chat history disabled, data is \textit{``retained for a time''}, leaving it susceptible to potential data breaches.

\subsubsection{AI-enabled Applications}
Conversational AI is increasingly being integrated into applications such as e-commerce and travel booking platforms. However, users who leverage these systems are concerned about the security of their products and services.

\shortsectionBf{Prompt Injections and Jailbreaks.}
Users worry about the susceptibility of their chat services to jailbreaks~\cite{li2023multi} and prompt injection attacks~\cite{perezignore}. They fear that these attacks could enable unauthorized access to sensitive data or allow users to bypass normal procedures in consumer-facing applications. For instance, one user wrote: \textit{``A lot of the companies with early integration are susceptible to prompt injections.''} Another user queried: \textit{``Say I wanted to make a chatbot for a business using the API, would I be able to prevent a user from feeding it prompts that would significantly change the output (like DAN for example)?''}

\shortsectionBf{Custom GPTs.}
The recent introduction of customizable GPTs~\cite{openai2024introducinggpts}, which users can configure with custom instructions, data, and API access and make available to other users via the GPT store~\cite{openai2024gptstore}, has also fueled concerns among users. Users are concerned that adversaries could manipulate these custom GPTs to expose private data (\eg custom instructions). For example, one user noted: \textit{``Even after telling my custom GPT to not leak the information from the knowledge database and the custom instructions through which it is trained, it's still revealing those things.''} Another user described how custom GPTs can be tricked into \textit{``spilling secrets''}, including \textit{``sensitive documents''}.

\subsubsection{Security of AI-generated code}

Although several users acknowledge the utility provided by many conversational AI platforms (\eg ChatGPT) that enable code generation, debugging, and even direct code execution through built-in interpreters, we find that many developers and professionals express concerns about security vulnerabilities present in the generated code, describing it as \textit{``bug filled''}. They worry that the code generated by these platforms does not \textit{``follow up to date security principles''}, and many recommend against its use in software systems that handle sensitive data. For example, one user shared: \textit{``[I am scared] at the idea of some startup using [AI-generated code] excessively for something that stores and handles sensitive data.''} Similarly, a security professional expressed his concern about the scale at which AI-generated code is deployed, stating that \textit{``Society is absolutely unprepared for the knock-on effects of releasing all this [insecure] code into production ... there are just not enough of us [security researchers] to handle this.''}

\shortsectionBf{Human vs. AI-Generated Code.}
Users argue that over-reliance on AI-generated code inadvertently leads to vulnerabilities in software systems. For example, one user commented: \textit{``This is how you get zero-day vulnerabilities in your microservice. Anyone relying on completely machine-generated code to save a few thousand dollars is a [fool].''} Interestingly, some users counter-argue that human code may contain similar vulnerabilities~(\eg \textit{``Hiring human programmers is also an excellent way to get vulnerabilities in your code, though.''})

\shortsectionBf{Persistent Vulnerabilities.}
Some users note that AI-generated code can perpetuate security bugs if they are common in the training data. For instance, one user described that even after a vulnerability in AI-generated code is spotted, the platform may continue to generate code with the same vulnerabilities due to its presence in the model's training data: \textit{``What happens when a hacker finds a vulnerability in AI-generated code but the [LLM] keeps re-creating the same vulnerabilities because the code is so common''}.

\subsection{Privacy Regulations Compliance}

Our analysis shows that many users demonstrate awareness regarding data and privacy protection laws, particularly the General Data Protection Regulation (GDPR)~\cite{gdpr} in the European Union (EU) and the Health Insurance Portability and Accountability Act (HIPAA)~\cite{hipaa} in the United States. 

\subsubsection{GDPR Compliance}
\label{subsubsec:GDPR} 

GDPR~\cite{gdpr} is an EU law that protects individuals' personal data, setting strict guidelines for data collection, storage, and user consent. Users are concerned that conversational AI platforms may not fully comply with GDPR requirements. For instance, one user flagged the presence of personally identifiable information (PII) in the AI platform's training data, stating, \textit{“Training data has PII, which breaks GDPR.”}

Under GDPR, individuals have a \textit{“right to be forgotten”}, which requires organizations to delete all user-associated data upon request \cite{gdpr}. Users express skepticism on conversational AI platforms' ability to meet this requirement, particularly given that user data might be embedded in the AI model itself (as discussed in Section~\ref{subsec:data_retention}). For instance, a user noted, \textit{``ChatGPT doesn't really forget data the same way as erasing under the EU right to be forgotten requires ... once a chat is used in machine learning, you can't ever truly erase it''}.

Users also discuss GDPR's restrictions on collecting data from minors and question whether AI platforms implement proper age controls to comply with these guidelines. For example, one user stated that there is \textit{``no age limit for minors''}. Another user wrote that the technology cannot \textit{``target''} children if it is to be compliant with GDPR.

\subsubsection{HIPAA Compliance}
HIPAA is a U.S. law that mandates the protection and confidential handling of sensitive patient health information by healthcare providers and associated entities~\cite{hipaa}. Remarkably, many users are not only aware of HIPAA's strict data protection requirements but also actively discuss how conversational AI platforms may fall short of these standards. Despite HIPAA's protections, users report instances they perceive as non-compliant. One user, for instance, shared that they have observed \textit{``multiple doctors''} using ChatGPT with sensitive medical data, raising alarm over potential HIPAA violations. Another user expressed concern over healthcare providers potentially uploading patient records to ChatGPT, remarking, \textit{``I really hope that you are not uploading others' medical records to ChatGPT. That would be a massive PII and HIPAA violation.''}

Users are similarly apprehensive of using conversational AI platforms in therapeutic contexts. They worry that, unlike interactions with licensed therapists, AI interactions are not governed by confidentiality rules (\eg \textit{``If you talk to a psychologist, there are rules ... [However,] everything you write to ChatGPT is free to use for retraining''}).

\subsection{Transparency and Control}

Users frequently report that the transparency and control measures in conversational AI platforms are inadequate or misleading, highlighting that the options provided are unclear, limited, or even manipulative.

\shortsectionBf{Data Controls.}
Many conversational AI platforms offer users some control over data sharing, including options to manage data usage for model training \cite{openaidatacontrols}. However, users report significant limitations and confusion surrounding these controls, which undermines their ability to manage their data.

Many users are concerned that data sharing for model training is typically enabled by default, requiring users to manually opt out. For instance, one user noted, \textit{``There's an option under settings on GPT-4 that's automatically checked yes for them to use your data for model training purposes.”} Users also discuss how data control options are unclear and challenging to navigate, which creates confusion about how to effectively disable data sharing for model training purposes. For example, one user stated that \textit{``users cannot easily opt out of data sharing''} in ChatGPT. Another user reported how the setting to opt out of sharing in ChatGPT was renamed, but none of the guides were updated accordingly: \textit{``It used to say `chat history and model training.' Now, it only says to `improve the model for everyone'.''}

\shortsectionBf{Dark Patterns.}
Many users believe that AI platforms employ dark patterns to subtly nudge them toward sharing data by limiting functionality or providing minimal privacy options. Users report functionality issues or software bugs when they opt out of data collection, which they feel pressures them into opting in. For instance, one user questioned if UI bugs (\eg broken scroll) were \textit{``intentional on OpenAI's part to cripple your experience if you don't want to share data for training?''}. Similarly, users wonder if opting out of data sharing prevented them from \textit{``getting access to the voice [chat] feature''} or \textit{``logging into [the ChatGPT] iOS app''}.

Users subscribing to paid plans for conversational AI platforms feel especially cheated and penalized for prioritizing privacy. For instance, one user noted, \textit{``I just paid for [ChatGPT Plus] and [I] can't believe that I have to keep chat history enabled to use [the] code interpreter, meaning I can't use any sensitive data for data analysis.''}

\shortsectionBf{Policy Changes and Consent.}
Users also express concerns about AI platforms' frequent and often confusing policy changes, especially when these updates affect default privacy settings or alter permissions without clear communication. Many users highlight inconsistencies in privacy information across official sources. For instance, one user stated \textit{``[ChatGPT's] website contains references to both the new and old [data] processes[,] making it hard to understand''}. Another user mentioned that the website lacks transparency and \textit{``clear notices''}, especially regarding whether users can fully opt out of data collection.

Some users discuss that policy changes may be designed to exploit default settings. For example, one user reported that a recent ChatGPT policy update switched their data-sharing preference to \textit{``yes''} even though they had previously opted out. Many users express frustration with being coerced into accepting new terms, as declining updates often prevent them from using their accounts altogether. For instance, one user noted, \textit{``If you do not agree to the updates, you may delete your account. But this sounds a bit too much like {`if you don't like this, you can go and f*** yourself'}.''}

\takeawaybox{Key Takeaways (RQ1).}
Our analysis highlights users' strong concerns around data privacy, security, and control in conversational AI platforms. Users are uneasy about extensive data collection, unclear data usage, and the permanence of stored information, particularly fearing inadequate protection of personal data. Despite transparency and control features, many find these insufficient or confusing, which intensifies mistrust. These concerns are further fueled by potential noncompliance with privacy and data regulations. 

\section{RQ2: Users' S\&P Attitudes}
\label{sec:attitudes}

Our analysis reveals that users can be categorized into four groups based on their S\&P attitudes (\ie behaviors and preferences) towards conversational AI platforms: \textit{cautious} $-$ aware of privacy risks and taking proactive steps to protect their data,  \textit{inquisitive} $-$ actively seeking information, \textit{privacy-dismissive} $-$ indifferent to privacy risks, and \textit{resigned} $-$ passively accepting privacy risks.

\subsection{Cautious Users}
Cautious users are highly aware of the privacy risks associated with conversational AI platforms. They prioritize data protection and actively implement measures to safeguard their information, often sharing tips and strategies with others. Their behaviors include adjusting privacy settings when using online AI platforms, opting for local solutions, and incorporating guardrails in AI-enabled applications.

\shortsectionBf{Securing Interactions with Online Platforms.}
To mitigate privacy risks from cloud-based conversational AI platforms, cautious users take several actions. They restrict sensitive data in their prompts by sanitizing inputs to remove personally identifiable information (PII). Some users manually scrub their input text before sending it to the AI, as one suggested: \textit{``You should scrub input text before sending it to the LLM.''} Others seek tools to automate this process or use hypothetical or anonymized data. For example, a user mentioned: \textit{``I also sanitize and anonymize private information by running the data through a Python script.''}

Many cautious users use data control options to disable training on their conversations. For instance, one user noted: \textit{``Turning off chat history and model training can be done to protect your privacy and ensure confidentiality.''} Others regularly delete chat histories or even their accounts to prevent data retention and distribution (\eg ``I do not want my private information shared with someone online.'').

\shortsectionBf{Using Local Solutions.}
Believing local solutions to be more secure and private than cloud-based ones, many cautious users prefer open-source LLMs (\eg Llama~\cite{llama2024}) that operate on their devices. For instance, one user stated: \textit{``An on-device AI is secure, more private, and provides more scope for personalization.''} Another asked for offline models to avoid data interception risks: \textit{``I worry that others can intercept my interaction ... is there a private, offline version?''} Users also report using local models in practice, removing reliance on third-party cloud infrastructure and maintaining direct control over their data. For instance, a user recommended: \textit{``If you want to use such generative models with sensitive data, I recommend that you use [an] open-source model like LLaMA or Mistral and deploy [it] on-premise.''}

\shortsectionBf{Implementing Guardrails.}
Cautious users extend their privacy-preserving practices to designing their own AI services. They stay informed about new attacks and implement guardrails to prevent exploitation. For example, one user noted: \textit{``I am creating a list of malicious prompts and different jailbreaking methods so it can identify when someone is trying to gain access.''} Another user designed a prompt to stop custom GPTs from sharing private instructions: \textit{``I've come up with a prompt which you can add to the end of your custom GPT instructions to protect it.''}

\subsection{Inquisitive Users}
Inquisitive users actively seek to understand the privacy and data management practices of conversational AI platforms. They frequently question specific functionalities, such as chat history and training settings, to assess how their data is used. These users often balance concerns about privacy with the desire for convenience and efficiency in their tasks.

\shortsectionBf{Engaging with the Community.}
Inquisitive users often turn to others in the community to gather information and perspectives on S\&P practices. They ask questions to make informed decisions, such as: \textit{``As OpenAI uses the data entered into GPT models to retrain them, would you be worried that the data you put in could be compromised if this data becomes exposed? Is this an issue that bothers you personally/with your work?''}, \textit{``If I'm paying for Copilot Pro, will Microsoft use [my data] to personalize ads and sell/transfer [it] to third parties?''}, or \textit{``Do you have chat history and training enabled?''}

\shortsectionBf{Consulting AI Agents.}
Some users directly ask AI agents about their security practices to gain insights. Some users trust the AI's responses. For instance, they asked the platform: \textit{``Is [ChatGPT's] chat history really deleted?''} and received an affirmative answer. Others remain skeptical, stating: \textit{``ChatGPT stores information about its users [but] it will prefer to reject this fact if you ask.''}

\subsection{Privacy-Dismissive Users}
These users exhibit indifference to privacy risks, prioritizing the benefits of AI platforms. They often criticize privacy regulations, which they perceive as barriers to technological advancement, and may dismiss others' S\&P concerns.

\shortsectionBf{Prioritizing Benefits Over Risks.}
These users believe the advantages of conversational AI outweigh potential privacy risks, adopting a pragmatic approach that values immediate utility. They highlight how existing devices are already conducting surveillance, \eg \textit{``We are exchanging modern convenience of technologies that make our lives easier for privacy. Every single device you buy or website you visit is already doing surveillance... why would I actually care?''}

\shortsectionBf{Criticizing Privacy Regulations.}
Privacy-dismissive users often advocate for more lenient privacy laws, viewing regulations like GDPR as overly restrictive and hindering innovation. For instance, one user noted: \textit{``The obsession with privacy in the EU is both hilarious ... Who cares if some random person overseas knows some of my personal data?''} They feel disappointed when privacy laws delay access to new features \eg ChatGPT's memory feature~\cite{openai2023memory}, which was delayed \textit{``because of privacy rights in the EU.''} Another user voiced frustration about not being allowed to waive privacy rights for early access: \textit{``As an EU citizen, I really hate that they don't even allow me to opt in.''}

\shortsectionBf{Dismissing Others' Concerns.}
Privacy-dismissive users may also overlook or trivialize others' concerns. For instance, one user responded to privacy worries by saying: \textit{``If you don't like it, don't use it. Who cares if they have your data? ... I'd give them all my data to improve the bot if I could!''} Similarly, they dismiss concerns about targeted advertising: \textit{``If it helps you change your thought patterns in a positive way, who cares what ads are tailored to you because of it?''}

\subsection{Resigned Users}
Resigned users feel overwhelmed by the pervasive nature of data collection and perceive resistance as futile. They believe that engaging with technology inevitably involves sacrificing privacy, expressing a sense of inevitability and helplessness regarding data privacy. For example, one user stated: \textit{``there is no running [from it]. [It's] scary to believe that the AI has the potential to access all this info about us as we use it.''} Another user noted: \textit{``At some point, I gave up. If I want the latest tech, I have to sell my soul/privacy.''}

These users argue that digital privacy is an illusion, as their data is collected regardless of their use of AI platforms. For example, one user noted: \textit{``[Your] data is being harvested no matter what you do if you have a phone so who cares''}

\takeawaybox{Key Takeaways (RQ2).}
Our analysis reveals four S\&P attitudes towards conversational AI platforms. Cautious users actively implement safeguards to protect their data, while inquisitive users seek to understand and navigate data usage policies. Privacy-dismissive users prioritize convenience, often criticizing strict privacy regulations, and resigned users feel that privacy is a lost cause in the digital age.

\section{RQ3: Longitudinal Analysis}
\label{sec:longitudinal_analysis}

\begin{figure*}[t!]
     \centering
     \begin{subfigure}[ht]{0.32\linewidth}
         \centering
         \includegraphics[width=\linewidth]{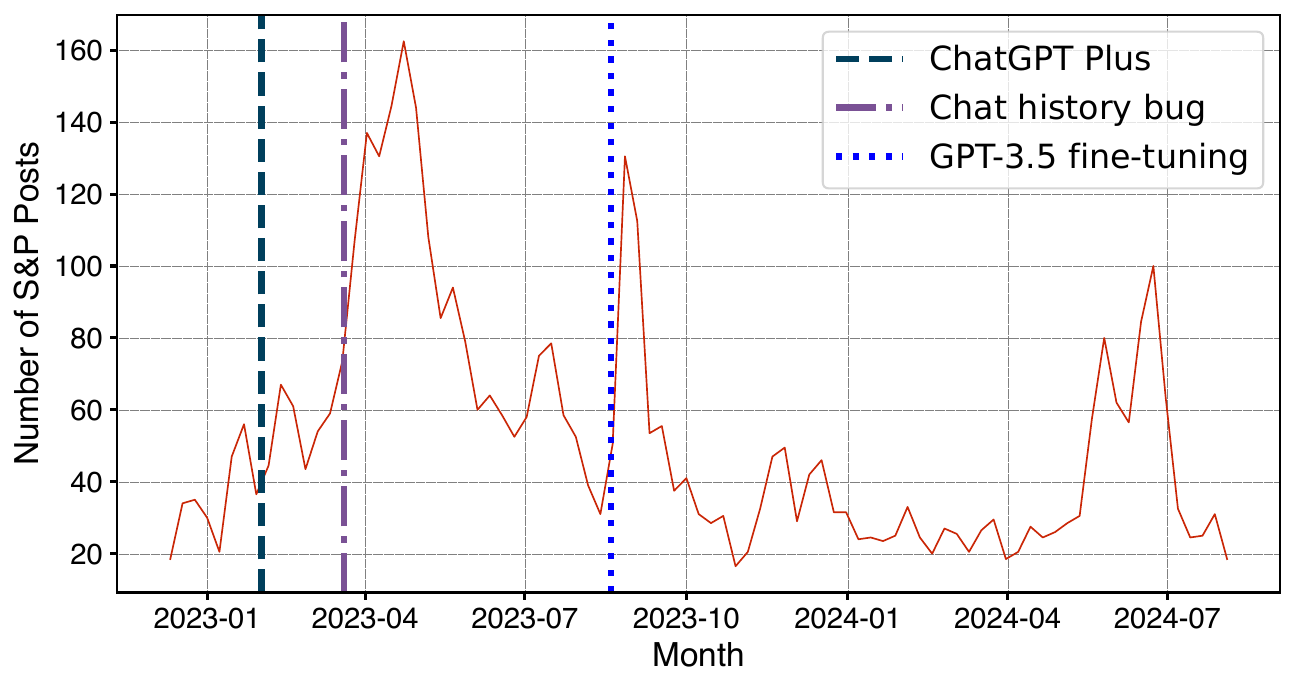}
         \caption{Personal data}
         \label{fig:personal_data}
     \end{subfigure}\hfill
     \begin{subfigure}[ht]{0.32\linewidth}
         \centering
         \includegraphics[width=\linewidth]{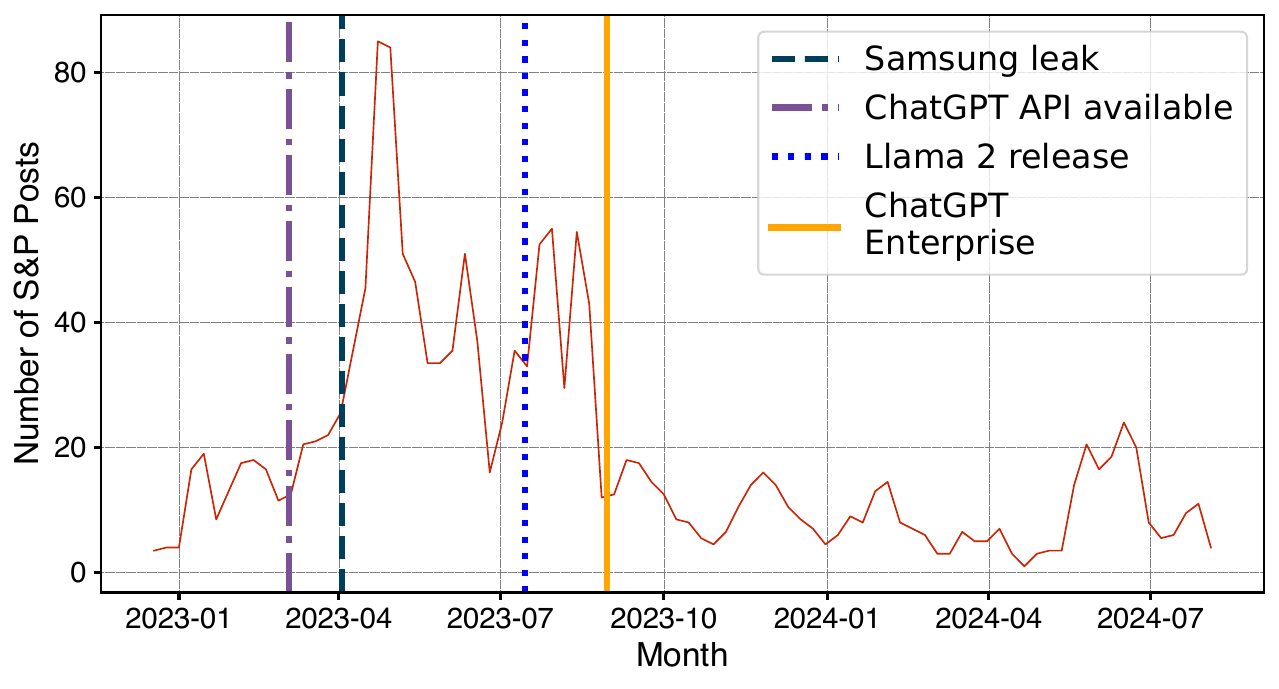}
         \caption{Proprietary information}
         \label{fig:proprietary_data}
     \end{subfigure}\hfill
     \begin{subfigure}[ht]{0.32\linewidth}
         \centering
         \includegraphics[width=\linewidth]{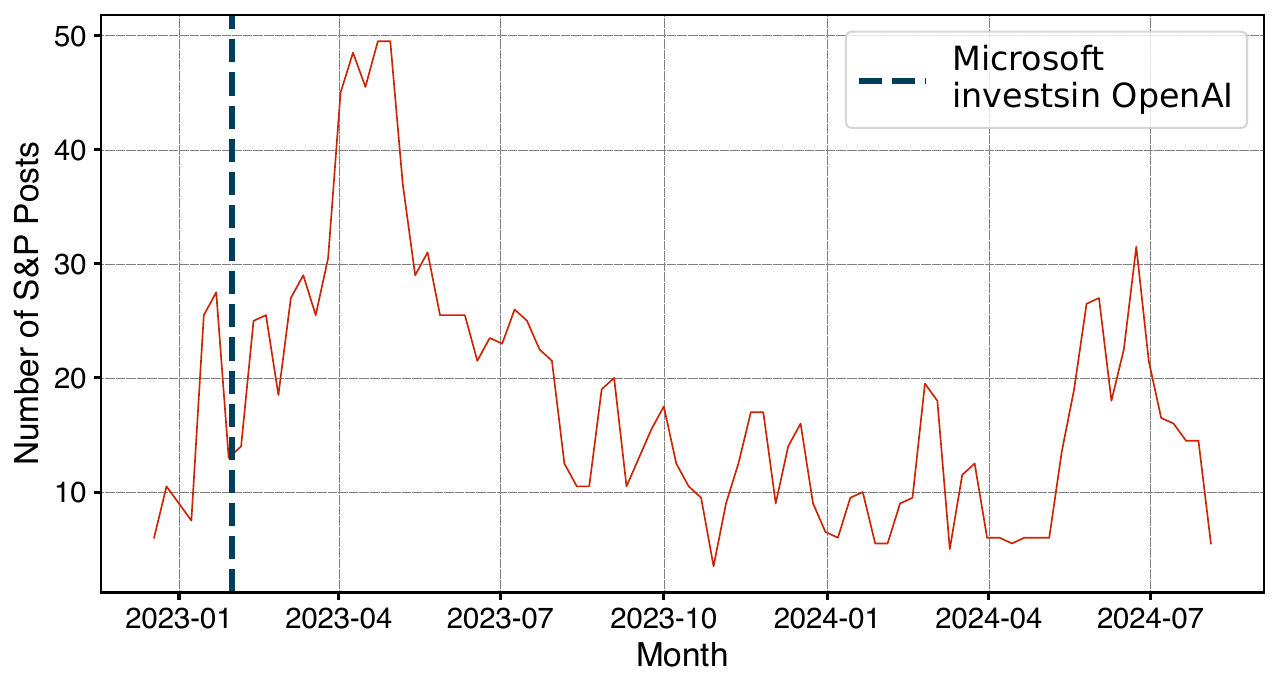}
         \caption{Third-party sharing}
         \label{fig:third_party_sharing}
     \end{subfigure}
     \begin{subfigure}[ht]{0.32\linewidth}
         \centering
         \includegraphics[width=\linewidth]{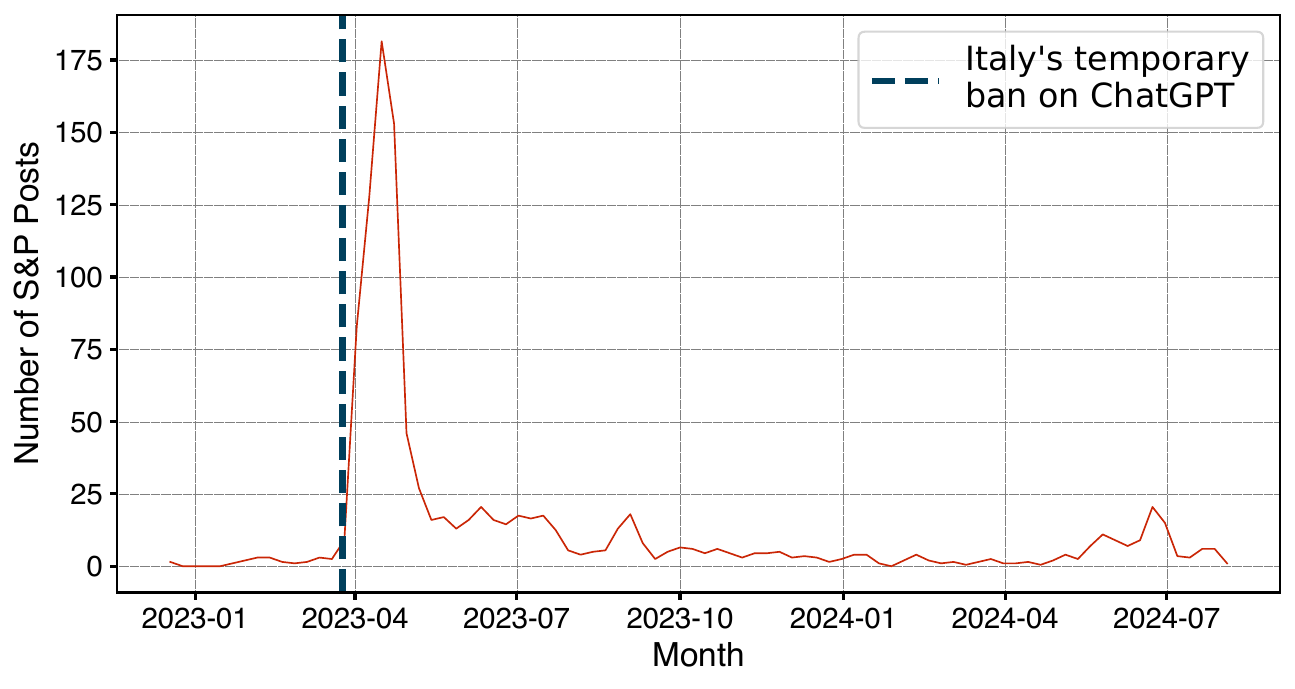}
         \caption{GDPR compliance}
         \label{fig:gdpr}
     \end{subfigure}\hfill
     \begin{subfigure}[ht]{0.32\linewidth}
         \centering
         \includegraphics[width=\linewidth]{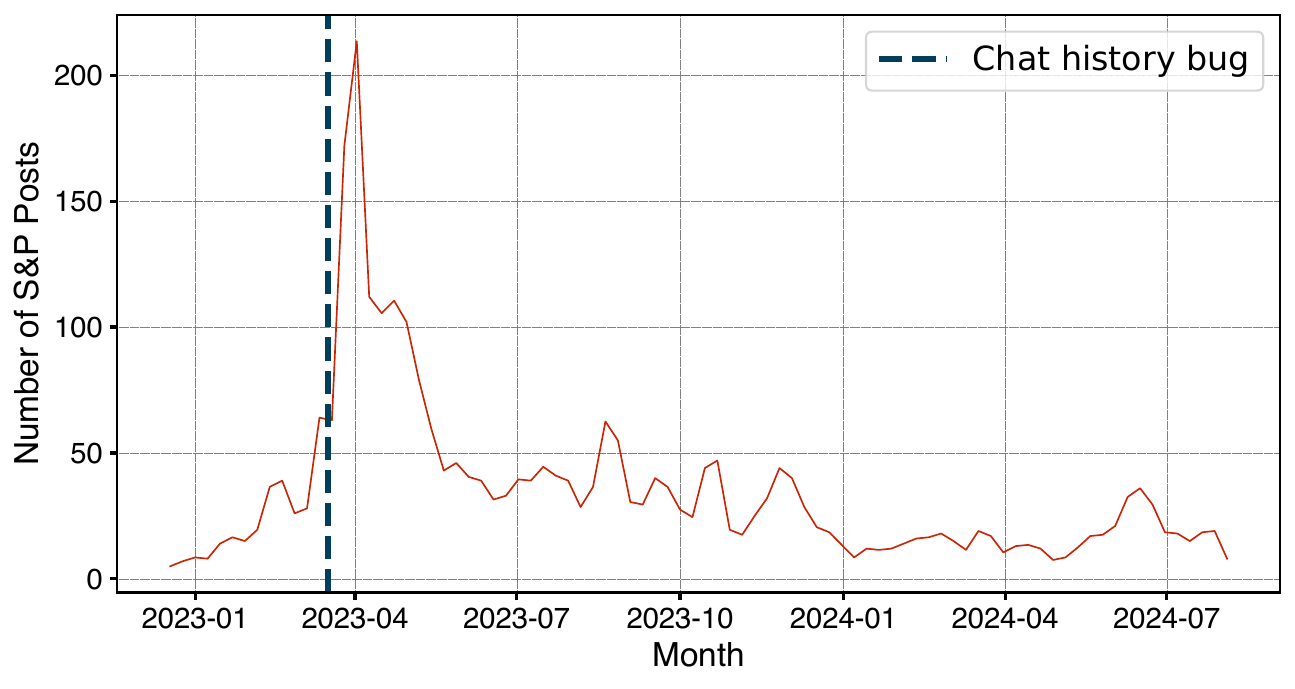}
         \caption{Platform security}
         \label{fig:platform_security}
     \end{subfigure}\hfill
     \begin{subfigure}[ht]{0.32\linewidth}
         \centering
         \includegraphics[width=\linewidth]{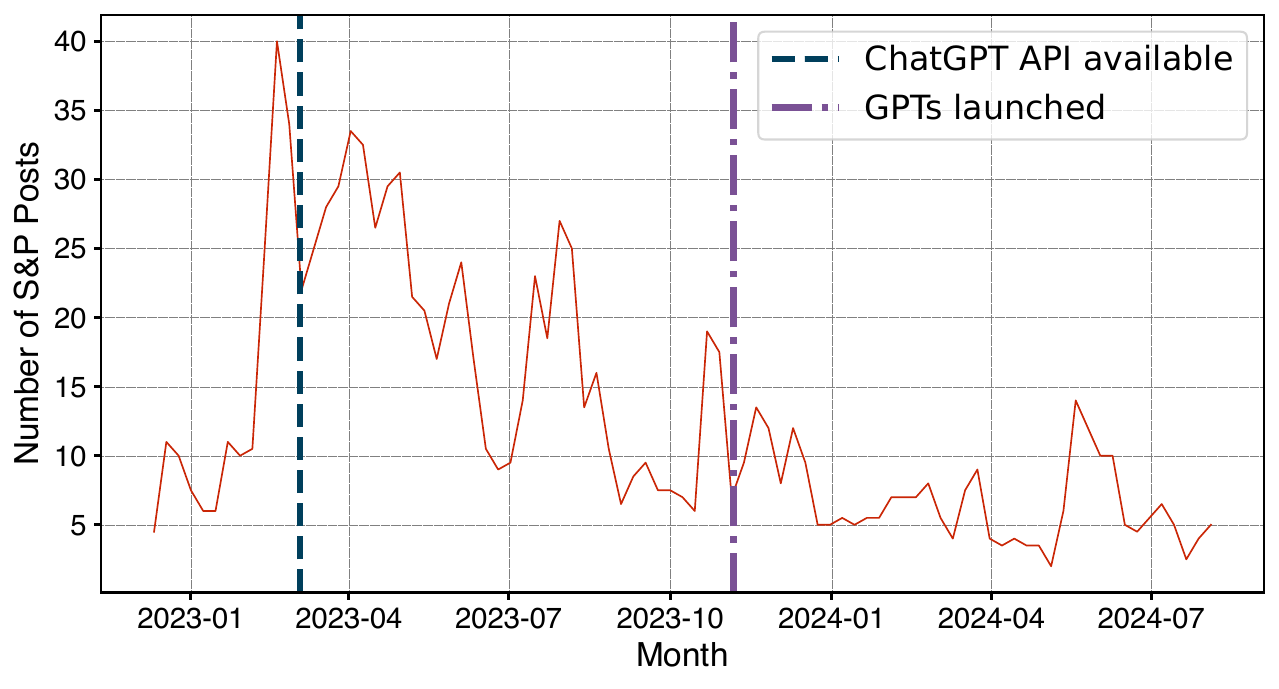}
         \caption{AI-enabled applications}
         \label{fig:llm_powered_apps}
     \end{subfigure}
    
     \caption{Impact of major events on S\&P discussions over time. We only display concerns that are significantly impacted (p < 0.05) by at least one event and annotate events that had a significant impact.}
     \label{fig:longitudinal_analysis}
\end{figure*}

To understand how users' S\&P concerns evolve over time and change in response to specific events, we first compiled a list of significant events related to conversational AI platforms. Two authors conducted a manual review of Google search results (from November $2022$ to July $2024$) using relevant keywords from our qualitative analysis (Section~\ref{sec:concerns} \& \ref{sec:attitudes}). By cross-referencing official announcements, we identified $38$ relevant events, including major announcements, platform and model launches, and feature releases. Our goal is not to compile an exhaustive list of events but to understand how different event types may impact different user concerns. Appendix~\ref{appendix:events} Figure~\ref{fig:events_timeline} shows the timeline of compiled events.

We conducted an interrupted time series regression analysis~\cite{mcdowall2019interrupted} to identify events significantly impacting S\&P discussions. We used an Ordinary Least Squares regression model to predict daily post counts within a seven-week window before and after each event. The model included time (in days), capturing the overall trend in number of discussions, an intervention term (binary variable: $0$ before and $1$ after the event, measuring immediate effects), and an interaction term (time x intervention) for post-event trend differences. We used a p-value of $0.05$ with Bonferroni correction~\cite{hochberg1987multiple} for multiple comparisons and manually validated the model's findings by inspecting discussions surrounding each event. Figure~\ref{fig:longitudinal_analysis} illustrates how major events influence S\&P discussions. 

\shortsectionBf{Launches and Acquisitions.}
We find that launch and acquisition events impact users' discussions about data collection and usage, specifically proprietary information and data sharing with third parties. For instance, when Microsoft invested $10$B dollars in OpenAI~\cite{ForbesMicrosoftChatGPT2023}, many users discussed the investment as an acquisition of OpenAI, considering Microsoft in control of OpenAI's data control policies. We also observe a peak in users' concerns about third-party data sharing (Figure \ref{fig:third_party_sharing}) after the event. Several users noted concerns about data monetization and targeted advertising. For instance, one user mentioned \textit{``Microsoft is a company which is hellbent on making money after spending money ... [They may implement] ads-based revenue like Google.''}

Another significant event that raised users' concerns about proprietary information collection was Meta's open-source release of LLama 2 in July $2023$~\cite{MetaLlama2_2023} (Figure~\ref{fig:proprietary_data}). Following the release, users discussed the privacy risks of using cloud-based conversational AI platforms and the potential privacy benefits of using open-source models. For example, a user noted \textit{``Instead of exposing potentially private code to a public service provider like OpenAI, why not use an open source local LLM (like Llama 2 or StarCoder)''}.

\shortsectionBf{Regulatory and Policy Changes.}
Italy banned ChatGPT in late March 2023 due to reported GDPR violations~\cite{nytimes2023chatgptitalyban}. We observe a peak in users' discussions about AI platforms' compliance with GDPR around this event (Figure \ref{fig:gdpr}). The event sparks user discussions around privacy. Some users note their preference for strict privacy laws supporting the ban and GDPR regulations (\eg \textit{``it's a matter of protection of personal data, how ChatGPT processes the user data is considered {`problematic'}''}). Others think that the privacy regulations are \textit{``stupid''} and excessive. For example, one user commented on the ban, saying \textit{``I'm so glad I don't live in Europe. I know we have a lot of bureaucracy in the United States, but it's on a whole [other] level over there.''} 

\shortsectionBf{Service and Feature Expansions.}
Our analysis shows that new services and features (\eg APIs for proprietary models and enterprise plans) influence discussions related to the collection of personal and proprietary information and the security of AI-enabled apps. We observe a peak in the concerns related to proprietary information (Figure \ref{fig:proprietary_data}) and the security of AI-enabled apps (Figure \ref{fig:llm_powered_apps}) in March 2023 when OpenAI enabled API access for their proprietary GPT models. Enterprises planning to incorporate these APIs in their operations have concerns about collecting and using sensitive corporate data. Users integrating these APIs in their products and services raise concerns about their susceptibility to exploits. We observe similar trends for proprietary data concerns with ChatGPT Enterprise's launch in late August 2023~\cite{openai2023chatgptenterprise} and for concerns about the security of AI-enabled apps with the release of Custom GPTs in early November 2023~\cite{openai2024introducinggpts}.

We also find that some service expansions impact users' discussions on personal data collection (Figure \ref{fig:personal_data}). For instance, the release of ChatGPT Plus in February 2023~\cite{openai2023chatgptplus} sparked discussions about whether the paid plan would ensure that user data would not be used for model training.

\shortsectionBf{Security Bugs and Breach Reports.}
We find that security incidents have a direct impact on user discussions related to S\&P. In March 2023, a bug in ChatGPT led to users seeing the titles of other users' conversations~\cite{Browne2023}. This correlated with an increase in concerns about personal data collection (Figure \ref{fig:personal_data}) and platform security (Figure \ref{fig:platform_security}). Users reported incidents like: \textit{``ChatGPT showed me another person's dialogues ... [It] doesn't open [the] chat when clicked, [but] still seems like a serious privacy problem.''} Another user questioned the platform's security \textit{``Has ChatGPT or me been hacked? [I've] never had these conversations.''}

A similar event occurred in April 2023 when Samsung employees reportedly leaked sensitive company data to ChatGPT~\cite{SamsungChatGPTLeak2023}, leading to the company eventually banning ChatGPT among employees~\cite{ray2023samsung}. We find that this event led to increased user discussions about proprietary information (Figure \ref{fig:proprietary_data}), primarily due to user concerns about the privacy of corporate or company data shared with the AI platform. Enterprises became more cautious, with some banning the use of ChatGPT among employees: \textit{``My company blocked ChatGPT... They are worried about security concerns.''}

\takeawaybox{Key Takeaways (RQ3).}
Our analysis shows that significant events, such as feature launches, regulatory changes, and security incidents, trigger distinct patterns in users' S\&P concerns. Key concerns are data sharing with third parties, corporate data privacy, and platform security, with notable spikes in concern following events like Microsoft's investment in OpenAI and the ChatGPT security bug.
\section{Discussion}

Our analysis of Reddit posts from the \texttt{r/ChatGPT} community provides valuable insights into users' security and privacy (S\&P) concerns toward conversational AI platforms. These findings reveal parallels with concerns about other emerging technologies while highlighting unique challenges posed by conversational AI platforms. In this section, we discuss the implications of our findings, situate them within the broader context of technology adoption and privacy concerns, and provide recommendations for key stakeholders.

\subsection{Key Insights and Implications}
\label{subsec:key_insights_and_implications}

Although users' concerns and attitudes toward conversational AI platforms mirror those of other computing platforms, such as smartphones and IoT devices, we identify unique factors inherent to conversational AI platforms that escalate or reshape these concerns. Moreover, we find that users' S\&P concerns evolve over time, influenced by platform updates, new features, and reported security incidents.

\subsubsection{Parallels with Other Technologies}
Our study shows how users worry about extensive data collection practices, echoing longstanding concerns in the S\&P community. Several of the S\&P concerns identified in our analysis parallel those found in other computing platforms and emerging technologies. For instance, concerns about personal data collection, such as PII, location data, and activity tracking, are not unique to AI platforms~\cite{agarwal2013no,seberger2021us}. Similarly, the unease over data sharing with third parties, the potential misuse of data for targeted advertising, and inadequate transparency and control mechanisms are common themes in S\&P research on devices such as smartphones~\cite{frik2022users} and IoT devices~\cite{zheng2018user} and platforms such as social media~\cite{rader2014awareness} and VR~\cite{abhinaya2025they}.

\subsubsection{Concerns Unique to Conversational AI Platforms}
Despite these similarities, conversational AI platforms present critical differences that exacerbate users' S\&P concerns, posing obstacles to secure adoption. First, the conversational format and LLMs' ability to emulate human empathy often lead users to disclose more sensitive or personal information than they would on other platforms (e.g., confidential business plans, proprietary code, or deeply personal struggles).

Second, conversational AI platforms use user data for model training, which can lead to the inadvertent memorization of sensitive information by AI models, a phenomenon distinct from traditional data-storage models. Memorized data can be regurgitated during inference, raising ethical and regulatory questions about a ``right to be forgotten''.

Lastly, conversational AI has seen an unprecedented and rapid scale of adoption. This widespread popularity has driven its integration into high-stakes sectors such as finance, healthcare, and enterprise services, where sensitive data and critical operations are involved. In these contexts, the natural language interface becomes a gateway to high-stakes data, creating novel security risks like prompt injections or jailbreaks that can expose secrets or allow unauthorized access to sensitive functionality.

\subsubsection{Validity of User Concerns}
As our analysis is based on user-posted content from Reddit, many concerns reflect users' perceptions of platform behavior. While these concerns are not always technically accurate, they offer valuable insights into user sentiment and potential trust or usability issues. We validate user concerns by analyzing platform documentation, feature updates, privacy policies (e.g., OpenAI's), and existing literature, categorizing them as presently valid concerns, resolved issues, or misconceptions.

\shortsectionBf{Presently Valid Concerns.}
We find that several concerns raised by users remain pressing and unresolved. For instance, LLM-generated code has been shown to contain security vulnerabilities, such as susceptibility to SQL injection and cross-site scripting attacks~\cite{pearce2022asleep}. Platforms remain vulnerable to prompt injections and jailbreaks, with studies showing how maliciously crafted inputs can override system instructions or expose sensitive information~\cite{liu2024formalizing, piltch2023chatgpt}. Concerns about memorization and lack of ``right to be forgotten'' also persist; prior work has demonstrated that LLMs can retain and regurgitate sensitive data from training sets, raising regulatory and ethical challenges~\cite{carlini2021extracting}. Lastly, model training and memory remain opted in by default on ChatGPT~\cite{openaidatacontrols}, which is a pre-selection dark pattern\cite{shi202550}. 

\shortsectionBf{Resolved Issues.}
Some concerns reflect actual incidents or bugs that platform developers have since addressed. For example, the ChatGPT bug that briefly exposed conversation titles to other users was acknowledged by OpenAI and is resolved~\cite{Browne2023}. Similarly, model training and chat history, which were initially coupled together~\cite{OpenAI2023DataManagement} (forced action dark pattern\cite{shi202550}), have now been decoupled~\cite{openaidatacontrols}.

\shortsectionBf{Misconceptions.}
Concerns about platforms accessing GPS location data without permission are unsubstantiated. We also found no discrepancies in features or access between accounts with and without model training disabled. These misconceptions imply that users have incorrect mental models of S\&P and conversational AI platforms' capabilities, indicating the need for platforms to educate users by making privacy-related information more accessible (Section~\ref{subsubsec:platform_recommendations}).

\subsubsection{Evolution of Users' Concerns}

Our findings indicate that users' S\&P concerns evolve over time, often in response to new features and reported incidents. Initially, many users express excitement about conversational AI platforms' capabilities but simultaneously exhibit caution due to underlying S\&P concerns. We find that the introduction of new functionalities (\eg custom GPTs), tends to reignite discussions about privacy, primarily due to a lack of transparency and detailed information about these features.

Moreover, users' concerns are exacerbated by the perceived inadequacy of transparency and control mechanisms. Confusion over privacy settings, policy changes, and data control options contributes to a sense of mistrust. Users often feel that their ability to manage their data is limited or obfuscated, leading to frustration and skepticism about the platforms' commitment to protecting their privacy.

Reported security incidents, such as data breaches or software bugs that expose user data, also significantly impact users' trust in the platforms. For example, the ChatGPT bug that revealed conversation titles to other users led to heightened concerns about data security and platform reliability. These incidents underscore the importance of robust security measures and transparent communication from platform providers to maintain user trust.

\vspace{-5pt}
\subsection{Recommendations}
\label{subsec:recommendations}

Based on our findings, we synthesize recommendations for stakeholders to address users' S\&P concerns and promote the safe, responsible use of conversational AI platforms.

\subsubsection{For Users}

Users play an active role in protecting their privacy and ensuring safe interactions through informed decisions and cautious behavior.

\shortsectionBf{Exercise Data Controls.}
Users should explore and use the privacy settings available on the platforms, such as opting out of data usage for model training when possible. Adjusting these settings helps limit the exposure of personal information and aligns with individual privacy preferences.

\shortsectionBf{Sanitize Inputs.}
To minimize the risk of unintentional data exposure, users should avoid sharing sensitive or personally identifiable information (PII) in prompts. Simple strategies (\eg replacing names, addresses, or other identifiers with placeholders) can provide an effective first layer of privacy protection. Prior work has demonstrated that prompt-level modifications, such as removing, masking, or substituting sensitive keywords, can substantially reduce privacy risks while preserving the utility of LLM output~\cite{zhu2024exploiting}.

\shortsectionBf{Read and Compare Privacy Information.}
Taking time to review platforms' privacy policies, especially simplified summaries when available, can help users make informed choices. Comparing data practices across platforms enables users to select services that align with their S\&P goals.

\subsubsection{For Platforms}
\label{subsubsec:platform_recommendations}
Conversational AI platforms bear significant responsibility in safeguarding user data and maintaining trust. To address users' concerns, platforms should enhance transparency, provide better control mechanisms, and prioritize user privacy in their operations.

\shortsectionBf{Improve Transparency.}
Platforms should prioritize clear and accessible communication regarding data collection, usage, and retention practices. Users often find privacy policies and terms of service documents dense and difficult to understand, leading to confusion and mistrust. To address this, platforms can take inspiration from the mobile ecosystem, where iOS and Android have introduced privacy labels~\cite{AppleAppPrivacyDetails} and data safety sections~\cite{GooglePlayIntegrityAPI} that simplify key information and make it more user-friendly. 

By adopting similar design patterns, highlighting data usage, model training purposes, and sharing practices, AI platforms can empower users to make informed decisions and compare privacy standards across services, ultimately enhancing trust and transparency.

\shortsectionBf{Design Better Data Controls and Nudges.}
Platforms can help alleviate concerns by offering robust and user-friendly data controls. Users have expressed frustration over the lack of clear settings to manage their data sharing preferences, especially regarding opting out of model training without losing essential features. Platforms can empower users to take control of their personal information by designing intuitive interfaces that allow them to easily adjust their privacy settings. Gentle interventions and nudges that direct users toward safer practices can help improve users' attitudes towards S\&P. Such proactive nudging strategies have shown promise in other domains. For instance, Wang~\etal~\cite{wang2013privacy} found that privacy nudges on Facebook could reduce unintended disclosures, while Li~\etal~\cite{SmartHome} advocate for privacy-focused nudges during smart home device setup.

\shortsectionBf{Educate Users.}
Prior research in S\&P has shown that even motivated users often struggle to make safe choices without clear, accessible guidance~\cite{whitten2005johnny}. Building on these insights, we recommend that platforms develop educational resources to help users understand their operations and data-handling practices. These resources (\eg tutorials, FAQ pages) can address common misconceptions, provide guidance on best practices for protecting personal data (\eg adjusting privacy settings), and explain how user data is used for model training. By providing clear, targeted education, platforms can foster a more informed user base.

\subsubsection{For Enterprises}

Organizations using conversational AI in their operations must proactively address S\&P concerns to protect both their interests and those of their stakeholders.

\shortsectionBf{Develop Clear Usage Guidelines.}
Developing clear guidelines for the use of AI platforms is essential. Enterprises should establish policies that outline acceptable practices for employees, particularly regarding the handling of sensitive or proprietary information. For instance, guidelines might prohibit the input of confidential company data into AI platforms unless certain privacy safeguards are in place. Training programs can educate employees about the risks and best practices, helping to prevent potential data leaks.

\shortsectionBf{Regular Security Assessments.}
Conducting thorough security assessments of AI-powered applications is another critical step. Enterprises should evaluate their systems for vulnerabilities, such as prompt injections or jailbreaks, which could compromise security or allow unauthorized access. Implementing technical safeguards, regular audits, and continuous monitoring can enhance the resilience of these applications against potential threats. Leveraging enterprise solutions offered by AI platforms can also address privacy and compliance requirements. Using these enterprise-grade services, organizations can ensure that their use of AI aligns with legal standards and protects sensitive information.

\subsubsection{For Policymakers}
Policymakers have a pivotal role in establishing frameworks that protect users while allowing technological innovation.

\shortsectionBf{Establish Clear Regulations.}
Policymakers should develop standards that define acceptable data practices for AI platforms, including requirements for transparency, user consent, and data protection. These regulations should specifically address issues such as data usage in model training, retention policies, and the handling of PII. For instance, laws could mandate that platforms obtain explicit consent before using user data for training purposes or sharing it with third parties.

\shortsectionBf{Promote Standardization of Privacy Information.}
By encouraging standardized privacy labels or data safety sections, policymakers can help users better understand and compare the data practices of different platforms. This standardization can drive industry-wide improvements in privacy standards and empower users to make informed choices.

\subsection{Limitations and Future Work}
\label{subsec:limitations}

Our study focuses exclusively on the \texttt{r/ChatGPT} subreddit, a leading discussion forum for conversational AI with a substantial user base. While this forum provides rich insights into user perspectives, we acknowledge our study's limitations. First, Reddit users tend to be more technologically savvy, and \texttt{r/ChatGPT} participants may possess greater knowledge about conversational AI platforms compared to the general public. While \texttt{r/ChatGPT} offers a large, active user base with a diverse range of perspectives, including discussions on alternative platforms and open-source models, it may not fully represent the broader population of conversational AI users.  Users who frequent Reddit or this specific subreddit may differ demographically or attitudinally from users engaging with conversational AI in other contexts (e.g., enterprise, non-English communities). These factors introduce a selection bias, potentially limiting the applicability of our findings to a broader audience.

Second, Reddit's pseudonymous nature limits access to demographic data such as profession, age, or location, making it difficult to contextualize user concerns or analyze trends across specific demographics. Future work will extend the analysis to other platforms such as Twitter, Discord, and community forums to capture a more diverse range of user perspectives and richer demographic or contextual data.

Third, while our S\&P-related posts classifier achieves high accuracy on our seed corpus, there may still be false positives or negatives affecting the reported prevalence of different themes. Automated classification may not fully capture the nuances of user concerns, potentially leading to overestimations or underestimations of certain issues. Moreover, some posts may include users quoting AI-generated content or users presenting AI-generated content as their own. Posts that are clearly AI-generated (\eg ``ChatGPT's views on humanity'') or unrelated to S\&P are labeled as not-S\&P during manual annotation and accordingly filtered by our S\&P classifier. Yet, distinguishing AI-generated text from user-written content remains an open challenge~\cite{wu2024detectrl}.

Given that our findings reflect specific concerns from professionals, developers, and other users using these platforms for specific applications, future work could explore users' S\&P concerns and attitudes in specific contexts and domains. By examining sector-specific issues, we can identify unique challenges and develop targeted recommendations for users and stakeholders in those fields. Similarly, future research should investigate the influence of geopolitical and cultural factors on user attitudes toward conversational AI platforms to understand regional differences and develop culturally sensitive solutions and policies.

\vspace{-8pt}
\section{Conclusion}
We conduct a large-scale analysis of online user posts to study users' S\&P concerns and attitudes toward conversational AI platforms. Our qualitative analysis shows that users are concerned with all stages of the data lifecycle—collection, usage, and retention—and seek better security, regulatory compliance, transparency, and control over their data. We also highlight how users' concerns evolve over time in response to major events. Based on our findings, we provide recommendations for different stakeholders involved.

\ifCLASSOPTIONcompsoc
  \section*{Acknowledgments}
\else
  \section*{Acknowledgment}
\fi

\noindent We thank our anonymous reviewers and shepherd for providing us with valuable feedback.
This work has been partially supported by UCI Academic Senate Council on Research, Computing and Libraries (CORCL) Award.
Any findings, conclusions, and recommendations expressed in this paper are those of the authors only.


\appendices
\label{section:appendix}

\section{Sub-reddits Related to Conversational AI}
\label{appendix:subreddits}
\noindent Table~\ref{tab:subreddits} shows the list of popular subreddits related to conversational AI. 
We explored these subreddits through manual review and keyword searches and found concerns and attitudes consistent with those observed in \texttt{r/ChatGPT}.

\begin{table}[h]
\centering
\caption{Top subreddits related to conversational AI, ranked by size.}

\label{tab:subreddits}
\setlength{\tabcolsep}{0.2em}
\resizebox{0.5\columnwidth}{!}{
\begin{threeparttable}
\begin{tabular}{l|c}
\hline
\textbf{Subreddit}  & \textbf{Members}      \\ \hline
r/ChatGPT           &   7.4M                \\        
r/OpenAI            &   1.9M                \\
r/CharacterAI       &   1.7M                \\
r/ChatGPTPro        &   271K                \\
r/LocalLLaMA        &   226K                \\
r/ChatGPTCoding     &   147K                \\
r/ClaudeAI          &   70K                 \\
r/Bard              &   44K                 \\ \hline 
\end{tabular}
\end{threeparttable}
}
\end{table}

\section{Keywords for Seed Corpus Creation}
\label{appendix:keywords}
\noindent Table~\ref{tab:keywords} shows the list of keywords for filtering posts. 

\section{S\&P Classification Results}
\label{appendix:sp_classification}
\noindent Table~\ref{tab:sp_classification} shows the performance of different models on S\&P classification task.

\begin{table}[h!]
\centering
\caption{S\&P Post classification results}

\label{tab:sp_classification}
\setlength{\tabcolsep}{0.45em}
\def\arraystretch{1.05}
\resizebox{\columnwidth}{!}{
\begin{threeparttable}
\begin{tabular}{l|c|c|c|c}
\hline
\textbf{Classifier}  & \textbf{Accuracy}    & \textbf{Precision}    & \textbf{Recall}   & \textbf{F1 Score}   \\ \hline \hline
DeBERTa\cite{he2020deberta}   & 0.97  & 0.91  & 0.76  & 0.83   \\
RoBERTa~\cite{liu2019roberta}       & 0.96  & 0.83  & 0.81  & 0.82   \\
GPT-4o~\cite{gpt4o}    & 0.91  & 0.71  & 0.92  & 0.80   \\
Gemini-1.5-Flash~\cite{gemini_flash}             & 0.76  & 0.45  & 0.95  & 0.61   \\ \hline \hline
\end{tabular}
\end{threeparttable}
}
\end{table}

\begin{table*}[t]
\centering
\caption{Keywords used to filter candidate posts for manual annotation. We applied regex-based matching to capture phrase variations (e.g., \texttt{r'delete(?:my|your)?data'} to match “delete my data” and “delete your data”). Certain terms like \texttt{access} were matched as standalone words to avoid unrelated results (e.g., \texttt{accessibility}).}
\label{tab:keywords}
\setlength{\tabcolsep}{0.2em}
\renewcommand{\arraystretch}{1.1}
\resizebox{0.9\textwidth}{!}{
\begin{threeparttable}
\begin{tabular}{p{4cm}|p{4cm}|p{9cm}}
\hline
\textbf{Group} & \textbf{Description} & \textbf{Keywords} \\ \hline \hline

General Keywords & Broad S\&P-related concerns. & privacy, security, permission, encryption, malicious, steal, access, secure, safe, confidential \\ \hline

Sensitive and Personal Data & Types of sensitive information users worry about. & personal/private/confidential/sensitive/corporate/company/client/ customer/health/medical/patient + info/data/doc/file, pii, personally identifiable info \\ \hline

Data Protection Laws & Legal or regulatory references. & gdpr, ccpa, hipaa, eu ai act, data regulation, data compliance, data protection \\ \hline

Storage and Deletion & Data deletion, retention, and storage practices. & data/account/chat/conversation + deletion/removal, data storage, data retention, delete + chat/conversation/account/data\\ \hline

Collection and Usage & How data is gathered, used, or shared. & data collection, surveillance, monitoring, data usage, data handling, data sharing, data selling, opt-out, privacy settings, data controls, disable/turn off + history/memory/training \\ \hline

Security Threats & System vulnerabilities and attacks. & prompt leak, prompt injection, jailbreak, guardrail, model attack, data theft, data + breach/leak/exposure/extraction \\ \hline

\end{tabular}
\end{threeparttable}
}
\end{table*}

\begin{figure*}[t]
    \centering
    \includegraphics[width=0.95\linewidth]{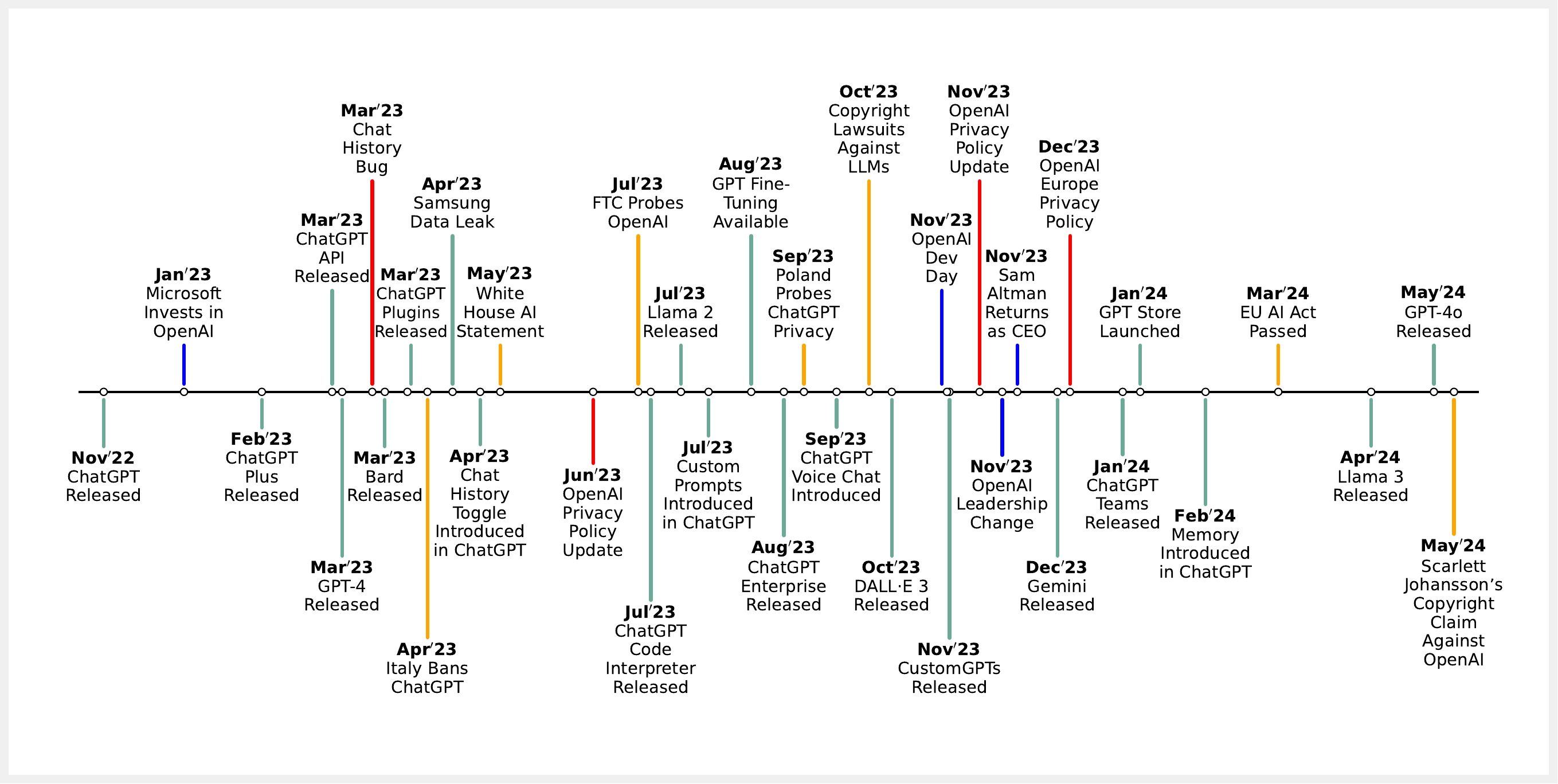}
    \caption{Key events in the development and adoption of conversational AI platforms, including product and feature releases (green), policy and regulatory changes (yellow), corporate developments (navy), and privacy and security events (red).}
    \label{fig:events_timeline}
\end{figure*}

\section{S\&P-Related Posts Over Time}
\label{appendix:sp_posts_over_time}
\begin{figure}[htbp]
    \centering
    \includegraphics[width=\linewidth]{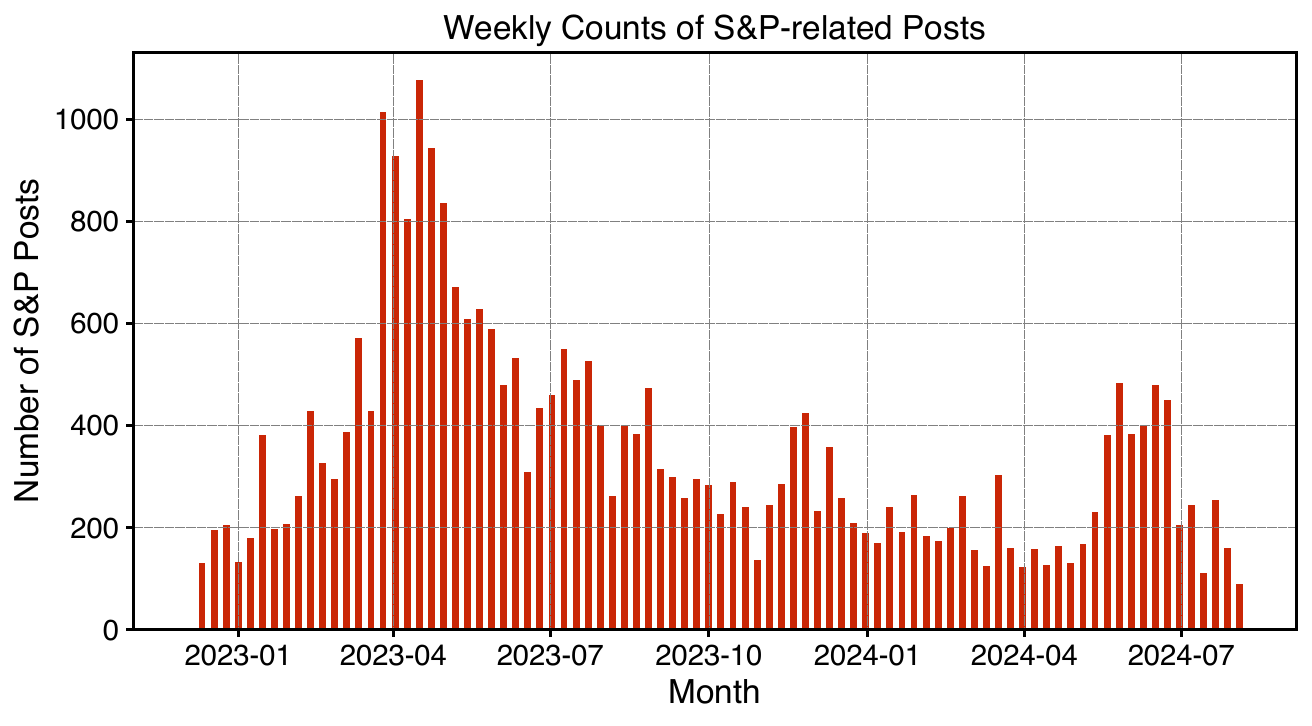}
    \caption{Weekly counts of S\&P-related posts.}
    \label{fig:sp_posts_over_time}
\end{figure}

\noindent Figure~\ref{fig:sp_posts_over_time} shows the weekly volume of S\&P-related posts over time, averaging 344 posts per week.

\section{Multi-Class Classification Results}
\label{appendix:multi-class}
\noindent Table~\ref{tab:multiclass_classification_results} shows the performance of the multi-class classification of S\&P-related posts using GPT-4o~\cite{gpt4o}.

\begin{table}[h!]
\centering
\caption{Performance of multi-class classification.}
\label{tab:multiclass_classification_results}

\setlength{\tabcolsep}{0.45em}
\def\arraystretch{1.05}
\resizebox{\columnwidth}{!}{
\begin{threeparttable}
\begin{tabular}{l|c|c|c|c}
\hline
\textbf{Type of Concern}  & \textbf{Accuracy}    & \textbf{Precision}    & \textbf{Recall}   & \textbf{F1 Score} \\ \hline \hline

\textbf{Data Collection}    & 0.94  & 1.0  & 0.89  & 0.94  \\
Personal Data    & 0.97  & 1.0  & 0.94  & 0.97  \\
Proprietary Information    & 0.96  & 1.0  & 0.81  & 0.90  \\ \hline

\textbf{Data Usage}       & 0.94  & 0.94  & 0.91  & 0.92   \\
Model Training    & 0.95  & 0.93  & 0.82  & 0.87  \\
Third-party Sharing    & 0.96  & 1.0  & 0.85  & 0.92  \\ \hline

\textbf{Data Retention}    & 0.97 & 0.83 & 1.0 & 0.91 \\ \hline

\textbf{Security Vulnerabilities}   & 0.96  & 0.87 & 0.93 & 0.90 \\ 
Platform Security    & 0.96  & 0.78 & 0.88 & 0.82  \\
LLM-powered Apps    & 1.0  & 1.0  & 1.0  & 1.0  \\
LLM-generated Code    & 0.99  & 0.8  & 1.0  & 0.89  \\ \hline

\textbf{Legal Compliance}    & 0.99  & 1.0  & 0.88  & 0.93  \\ 
GDPR   & 0.99  & 1.0  & 0.8  & 0.89  \\
HIPAA    & 1.0  & 1.0  & 1.0  & 1.0  \\ \hline

\textbf{Transparency \& Control}  & 0.95  & 0.71  & 1.0  & 0.83  \\  \hline

\end{tabular}
\end{threeparttable}
}
\end{table}

\section{Events Related to Conversational AI Ecosystem}
\label{appendix:events}
\noindent Figure~\ref{fig:events_timeline} shows the timeline of major events related to conversational AI platforms.

\end{document}